\documentclass[a4paper,11pt]{article}
\makeatletter
\def\@fpheader{\relax}
\makeatother

\usepackage{jheppub} 
\usepackage{background}

\usepackage{lineno}
\usepackage{xspace}

\usepackage{float}
\usepackage{comment}
\graphicspath{{figures/}}

\newcommand{\pf}{\ensuremath{\text{PREFACE}}\xspace}
\newcommand{\f}{\ensuremath{\text{FACET}}\xspace}

\arxivnumber{} 

\title{\boldmath PREFACE\footnote{\textbf{P}ioneer \textbf{R}are-\textbf{E}vent \textbf{F}orward \textbf{A}pparatus for \textbf{C}ollider \textbf{E}xperiments}: A search for long-lived particles at the Large Hadron Collider}

\backgroundsetup{contents={}}

\author[a,*]{B. Hacisahinoglu,}
\author[b]{S. Ozkorucuklu,}
\author[c]{M. Ovchynnikov,}
\author[d]{M.G. Albrow,}
\author[e]{A. Penzo}
\author[f]{\\and O. Aydilek}

\affiliation[a]{Institute of Graduate Studies in Sciences, Istanbul University, 34116, Istanbul, Turkey}
\affiliation[b]{Department of Physics, Istanbul University, 34134, Istanbul, Turkey}
\affiliation[c]{Theoretical Physics Department, CERN, 1211 Geneva 23, Switzerland}
\affiliation[d]{Fermilab, Chicago, USA}
\affiliation[e]{Department of Physics and Astronomy, The University of Iowa, Iowa City, USA}
\affiliation[f]{Department of Physics, Erzincan Binali Yıldırım University, 24002, Erzincan, Turkey}
\affiliation[*]{Corresponding author}

\emailAdd{Burak.Hacisahinoglu@cern.ch, Suat.Ozkorucuklu@istanbul.edu.tr, Maksym.Ovchynnikov@cern.ch, Albrow@fnal.gov, Aldo.Penzo@cern.ch, Orhan.Aydilek@erzincan.edu.tr}

\abstract{
The Standard Model (SM) fails to explain many problems (neutrino masses, dark matter, and matter-antimatter asymmetry, among others) that may be resolved with new particles beyond the Standard Model (BSM). The fact that such new particles are not yet observed may be explained either by their too high mass or by very small coupling to SM particles. The latter case implies long lifetimes. Such long-lived particles (LLPs) would have signatures different from those of SM particles. Searches in the ``central region" are covered by the LHC general purpose experiments. The forward small angle region far from the interaction point (IP) is unexplored. Such particles would have large energy $E$ = $\mathcal{O}$(1~TeV) and Lorentz time dilation factor $\gamma = E/m \approx 10^2 - 10^{3}$, hence long decay distances. A new class of specialized LHC detectors dedicated to LLP searches has been proposed for the forward regions. Among these experiments, FASER is already operational, and FACET is under consideration at a location 100 meters from IP5 (the CMS intersection). However, some features of FACET require a specially enlarged beam pipe, which cannot be implemented for Run 4. In this study, we explore a simplified version of the proposed detector, \pf, compatible with the standard LHC beam pipe in the HL-LHC in Run 4. Realistic \textsc{Geant4} simulations are performed and the background is evaluated. A preliminary analysis of the physics potential with the \pf geometry indicates that several significant channels could be accessible with sensitivities comparable to FACET and other LLP searches.}


\arxivnumber{2502.14598}

\begin{document}
\maketitle
\flushbottom

\section{Introduction}

Despite its success the Standard Model (SM) of elementary particles cannot address several crucial questions — such as neutrino masses, dark matter, and matter-antimatter asymmetry — that may require a new theory and new particles beyond the Standard Model (BSM). Experiments at the LHC are making significant efforts to search for BSM phenomena and identify their possible constituents \cite{Alimena:2019zri,BAWA2015277,Policicchio:2015spa}. In the near future the LHC will operate at even higher luminosities \cite{Aberle:2749422,Brüning2019}, enabling a deeper exploration of these questions \cite{Dainese:2703572}. This is particularly important for discovering new particles that may have extremely small cross sections or weak couplings to SM particles \cite{SHCHUTSKA2016656,RUHR2016625}. 

\subsection{BSM benchmark models of a dark sector}

A relatively simple, coherent, and predictive framework \cite{Fischer2022, Mitsou:2017ssh} for BSM states is often linked to a so-called ``dark sector," which interacts with the SM ``visible" sector via ``portals"~\cite{Alekhin:2015byh,Beacham:2019nyx,Antel:2023hkf}. Due to the constraints imposed by SM gauge symmetry and fields, only a few viable portal options exist: vector, scalar, fermion, and axion-like particle (ALP) portals. These straightforward and well-motivated benchmark models predict a wide range of phenomena, driving the development of many new searches and specialized experiments that complement other (e.g. cosmic) probes.

Depending on the model, a long-lived particle (LLP) may have one or two couplings governing both the production channels of the mediator or rare SM particle and its visible decay modes into SM final states \cite{Batell:2022dpx, Lu2023}.

\subsection{Proposed long-lived particle search experiments}
Past experiments have only excluded large couplings of the portal particles to new physics. Since their lifetime is inversely proportional to the squared coupling, they have relatively long lifetimes and are classified as LLPs. 

There are several reasons why LLPs, if they are produced at the LHC, might have escaped detection, including the smallness of the event rate (proportional to the squared coupling times the LLP decay probability, which is typically tiny), absence of efficient selective triggers, and overwhelming SM backgrounds.
To address these issues, dedicated experiments, pioneered by FASER \cite{Abreu_2024}, have been proposed to search for LLPs in the regions forward of the LHC interaction points \cite{Feng_2023}. While the solid angle $\Delta\Omega$ covered by these experiments is generally small, they have a large acceptance in pseudorapidity $\eta$ and azimuth $\phi$. The solid angle distributions of the LLP fluxes are maximal at large $\eta$, which partially compensates for the smallness of the solid angle coverage.

\section{Long-lived particle search at LSS5}
The FACET (Forward-Aperture CMS ExTension) detector, proposed \cite{Cerci:2021nlb} for installation at the IP5 intersection (CMS), stands out for its relative proximity to IP5 ($z \geq$ 100 m) and a fiducial volume four times longer than that of FASER, resulting in a larger acceptance. Another notable feature is its enlarged beam pipe, with a radius of r = 0.5 m between z = 100 m and z = 120 m ensuring LHC-quality vacuum conditions within the decay volume. The FACET proposal and its associated physics case has generated significant interest in the search for LLP production in the forward direction. Comparative studies with FASER and other proposed LLP detectors indicate that it performs favorably across numerous benchmarks \cite{Beacham:2019nyx}. Although initially aimed at the first HL-LHC Run 4, the special enlarged beam pipe is not possible for that run. To be compatible with the HL-LHC Run 4 beam pipe, a simplified version of FACET was considered \cite{Penzo2023}; it is described here as PREFACE (Pioneer Rare-Event Forward Apparatus for Collider Experiments)  and its physics case is discussed.

Preliminary studies of the physics potential of the PREFACE geometry suggest that several important channels are accessible with sensitivity comparable to, or exceeding, that of FACET and FASER.
FASER is installed in a shielded side tunnel of LHC, far from the nearest interaction region (ATLAS), while FACET and PREFACE are located in the LHC tunnel at only about 100 m from the CMS interaction point IP5 and are designed to sustain  a high radiation environment. 
A "near location" downstream the TAXN, has been considered~\cite{Feng:2017uoz} for FASER, but has large backgrounds. A location upstream the TAXN, described here for FACET or PREFACE, is preferable for a more robust detector assembly; in reference~\cite{Feng:2017uoz} a similar option is outlined in Appendix A.

The new PREFACE geometry, installed only above the standard LHC beam pipe, avoids the very high fluxes of charged particles swept to left and right by the beam separation dipole D1. Backgrounds from particles showering in the beam pipe are significantly reduced by a shielding plate between the pipe and the detectors. Reconstructing tracks in real time, with AI techniques, allows rejection at the trigger level of residual background from interactions on the beam pipes and inside the PREFACE setup, including on the air in the decay volume. Most interactions of fast neutrons and $K^0$ in air can be rejected by topology.

\subsection{Experimental conditions at LSS5}

\begin{figure}[h]
    \centering
    \includegraphics[width=0.6\linewidth]{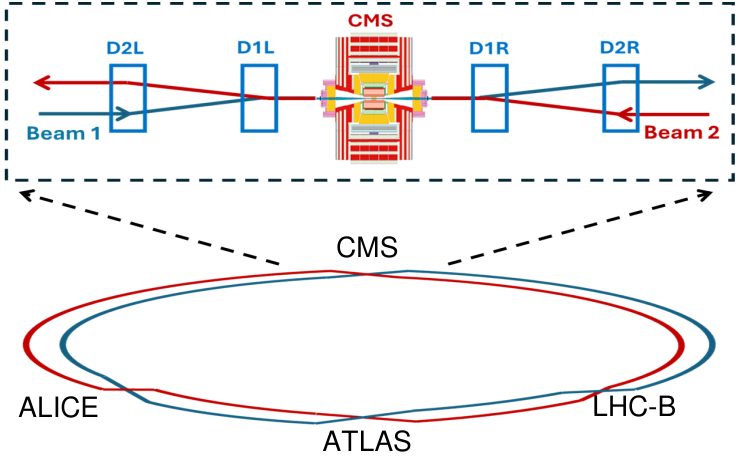}
    \caption{A sketch of the LHC ring, with the four major experiments studying pp, pA and AA collisions (usually A is Pb but lighter nuclei are possible).}
    \label{fig:LHC-ring}
\end{figure}

Figure~\ref{fig:LHC-ring} gives a schematic representation of the LHC beams colliding at the experiments, with the region around IP5 expanded to show the beam trajectories and lattice elements. The two beams are bent into the same straight section where they collide, using insertion stages made of dipole magnets, D1 and D2, which bend the incoming beams into the collision point, and bend the outgoing beams back into separate pipes at the TAXN absorber.

The intersecting of the beams is further complicated by a small inclination (typically 150 $\mu$rad) of the beams, either in the vertical or horizontal planes \cite{Arduini:2016xsb}. In each bunch crossing over 100 pp collisions may occur, each producing a large number of particles in the full solid angle (4$\pi$ steradians). Detectors surrounding the IP measure most of the reaction products, but many escape detection through the vacuum pipes which cannot be equipped with detectors. Two exceptions are  Zero Degree Calorimeters (ZDC)~\cite{Grachov:2010th} between the beam pipes for neutral particles and detectors in ``Roman Pots" for diffractively scattered protons without disturbing the beams~\cite{Albrow:1753795}. However, many particles hit the beam pipe and interact creating showers of secondary particles which can be a background for detectors located nearby.

\begin{figure}[h]
    \centering
    \includegraphics[width=1. \linewidth]{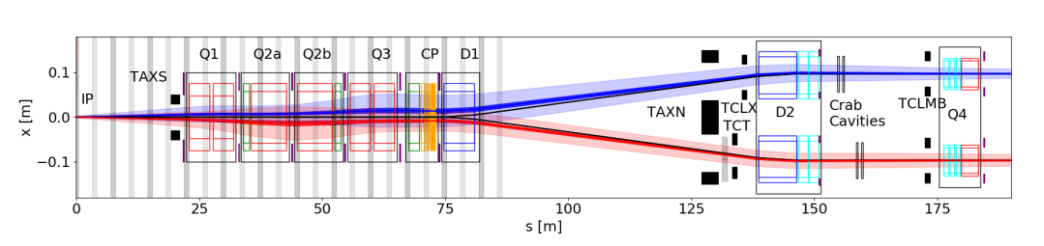}
    \caption {The overall layout downstream of the Point 5 interaction region (schematic). The incoming (blue) and outgoing (red) beams are in a common beam pipe between the beam separation dipole D1 and the TAXN. The light regions show the 14.2 $\sigma$ beam envelope~\cite{Arduini:2016xsb}.}
    \label{fig:CMS_downstream}
\end{figure}

\subsubsection{FACET overview}

\begin{figure}[h]
    \centering
    \includegraphics[width=1\linewidth]{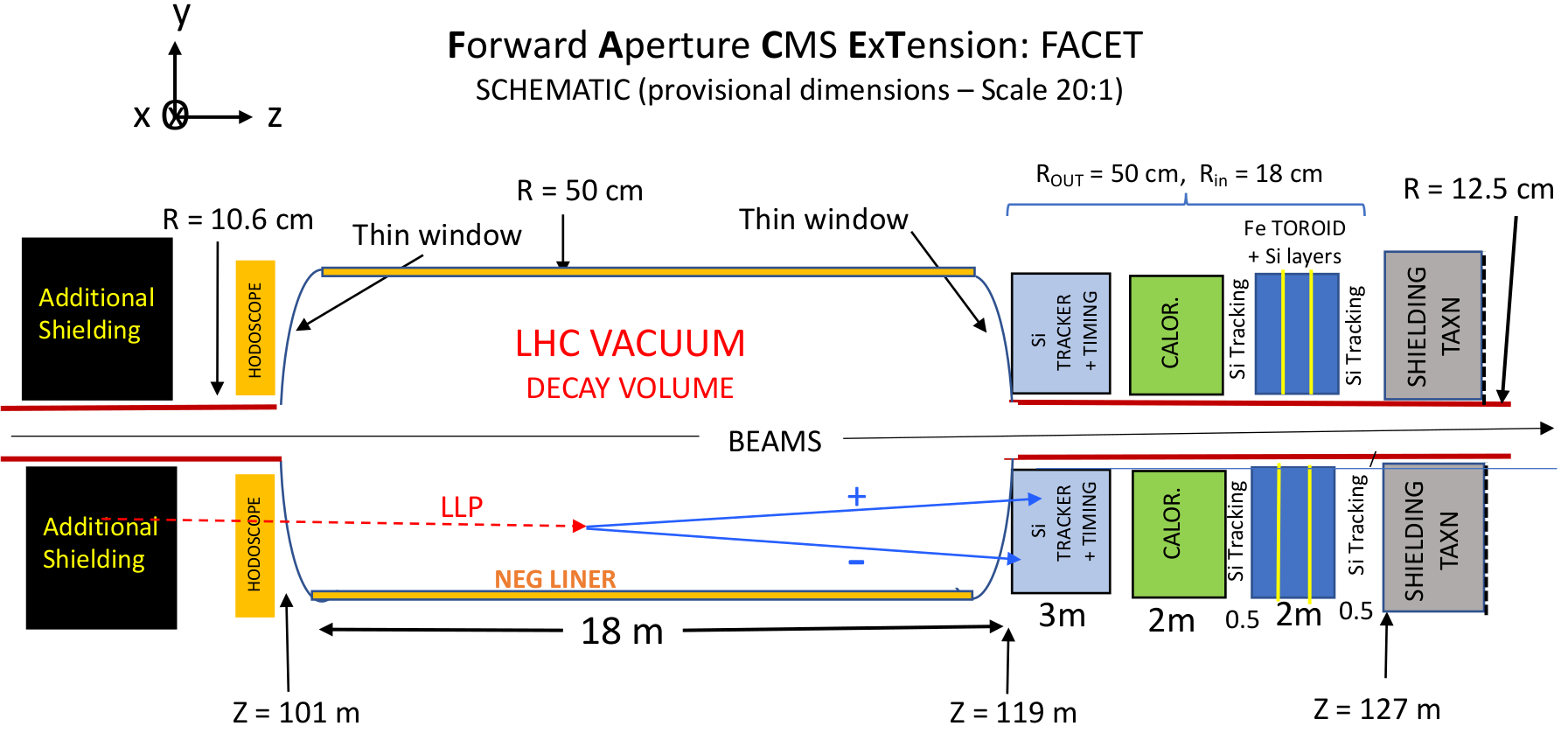}
    \caption{Schematic layout of the proposed FACET spectrometer.}
    \label{fig:schema_FACET}
\end{figure}

The location foreseen for FACET in LSS5 (figure~\ref{fig:CMS_downstream}) is in the field-free region between the 35 T$\cdot$m superconducting beam separation dipole D1 at z = 80 m and the TAXN absorber at z = 128 m. In figure~\ref{fig:schema_FACET}, the setup is shown schematically. The decay volume is an enlarged beam pipe with radius R = 0.5 m between z = 101 m and z = 119 m; the detectors following the decay volume have full azimuthal coverage with annular shape of inner radius $\text{R}_\text{in}=18$ cm and outer radius $\text{R}_\text{out}=50$ cm, covering polar angles $1.5<\theta<4$ mrad. The background is greatly reduced because of $200-300\,\lambda_{int}$ of magnetized iron in the LHC quadrupole magnets Q1-Q3 and dipole D1, and LHC quality vacuum ($10^{-7}$ Pa - $10^{-9}$ Pa) in the decay volume.

In order to estimate the backgrounds detailed \textsc{Geant4}~\cite{AGOSTINELLI2003250} simulations were performed with particles produced from a \textsc{Fluka}~\cite{fluka2024} simulation of CMS interactions at $\sqrt{s}=14$ TeV, giving particle distributions at z $\approx$ 100 m. \textsc{Geant4} simulation then tracks the particles produced in \textsc{Fluka} through the detectors.

\subsubsection{Backgrounds in FACET}
\begin{figure}[h]
    \centering
    \includegraphics[width=0.48\linewidth]{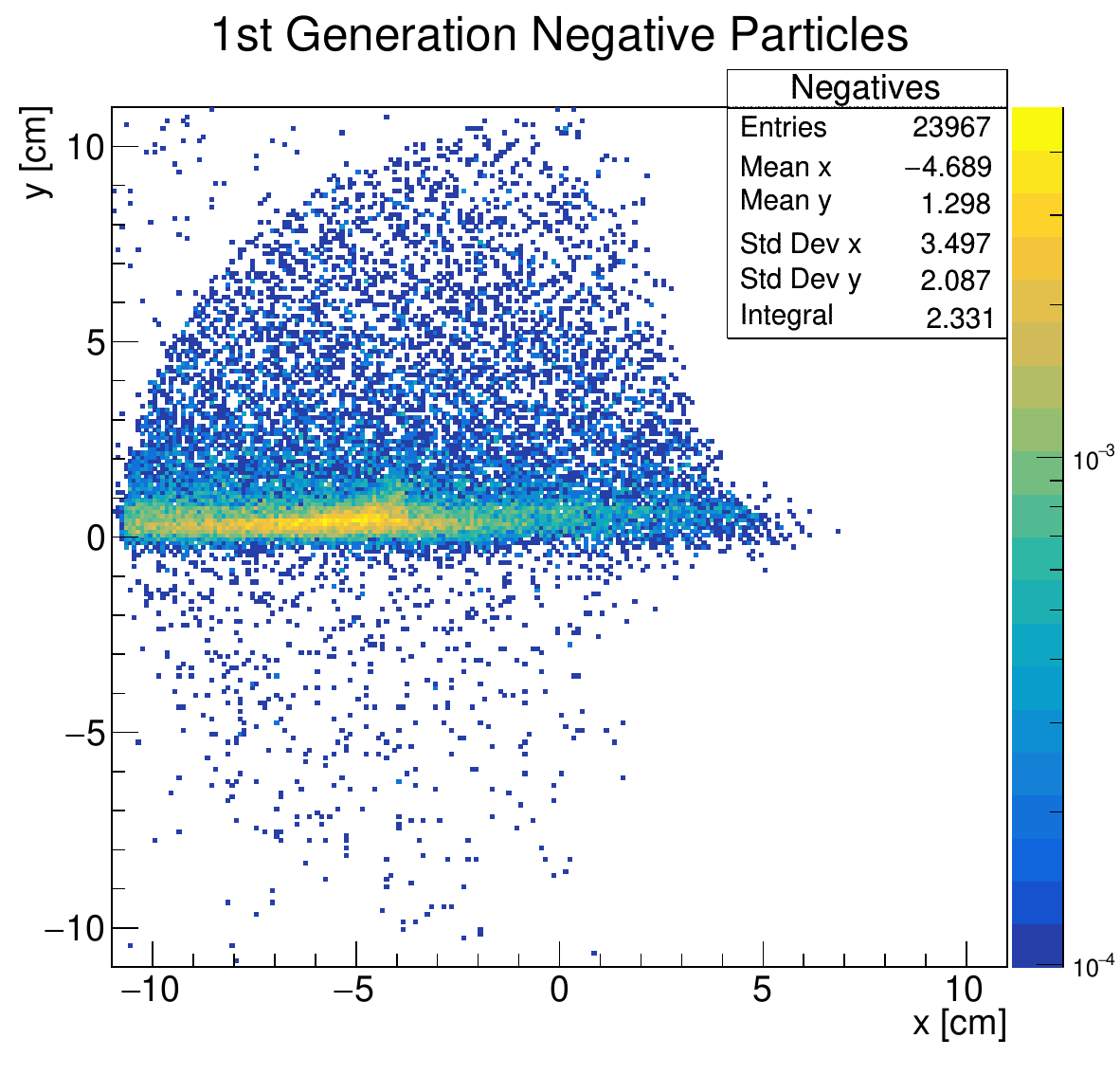}
    \includegraphics[width=0.48\linewidth]{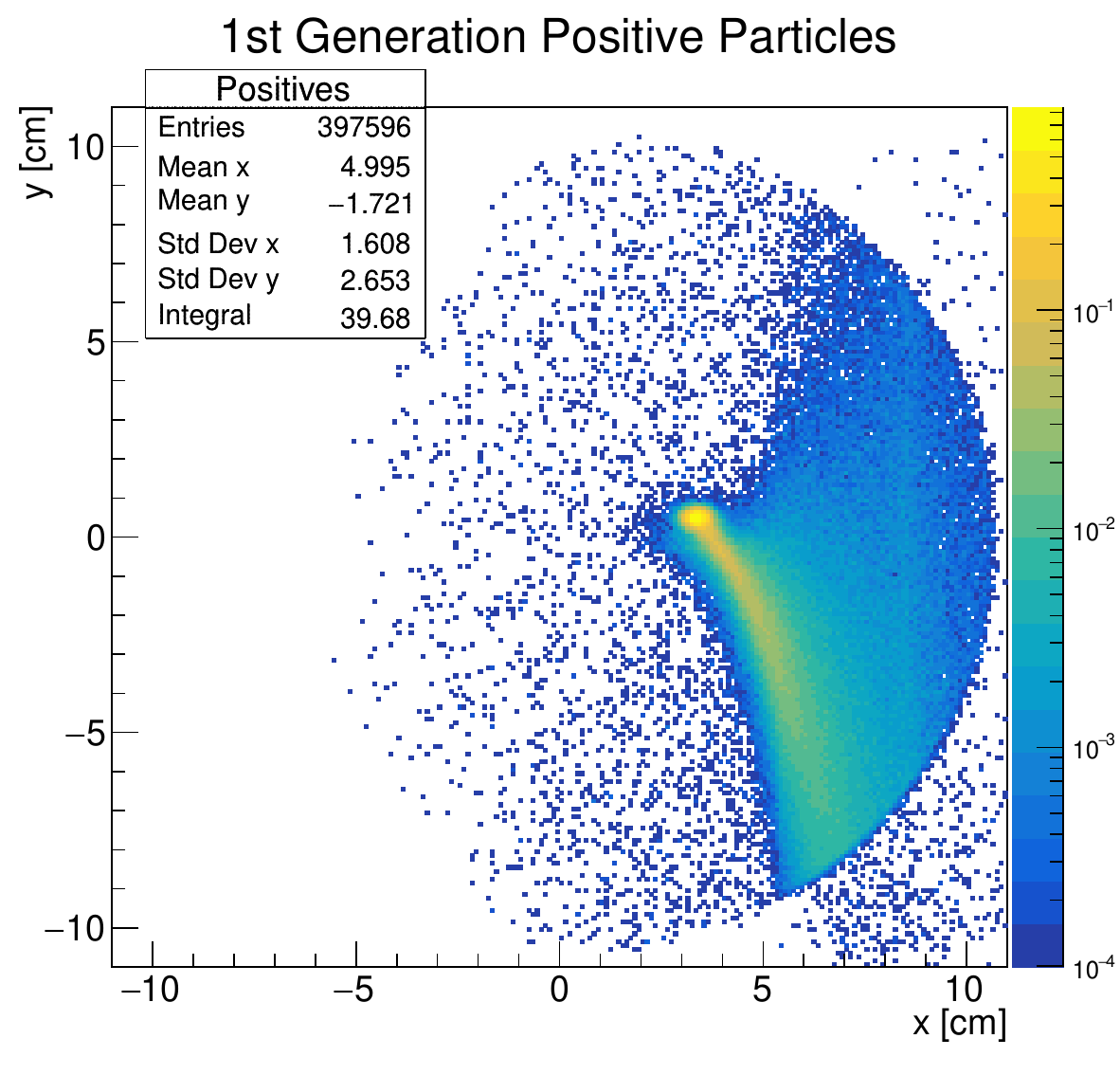}
    \caption{The distribution in the transverse plane of negative (left) and positive (right) particles at z = 100 m for particles originating at the interaction vertex at IP5. Colors show the flux in (mm)$^{-2}$ for 140 pp collisions.}
    \label{fig:ch_neg_and_pos_at_100m}
\end{figure}

The D1 magnet bends charged particles in the horizontal plane, positives to the right and negatives to the left, see figure~\ref{fig:ch_neg_and_pos_at_100m}, where those hitting the beam pipe produce the most background. The spot at x = 3 cm, y = 0 cm shows diffractively scattered protons with energy $\sim$~7~TeV, those with lower energy form the tail, downward because of the vertical crossing angle of the beams. (For about half of the run the crossing angle will be changed so the high intensity tail will be upward.) The negative particles from IP5, mostly $\pi^-$,   are concentrated in the horizontal plane to the left of the central beamline, and most have momenta of a few TeV.

\begin{figure}[h]
    \centering
    \includegraphics[width=0.49\linewidth]{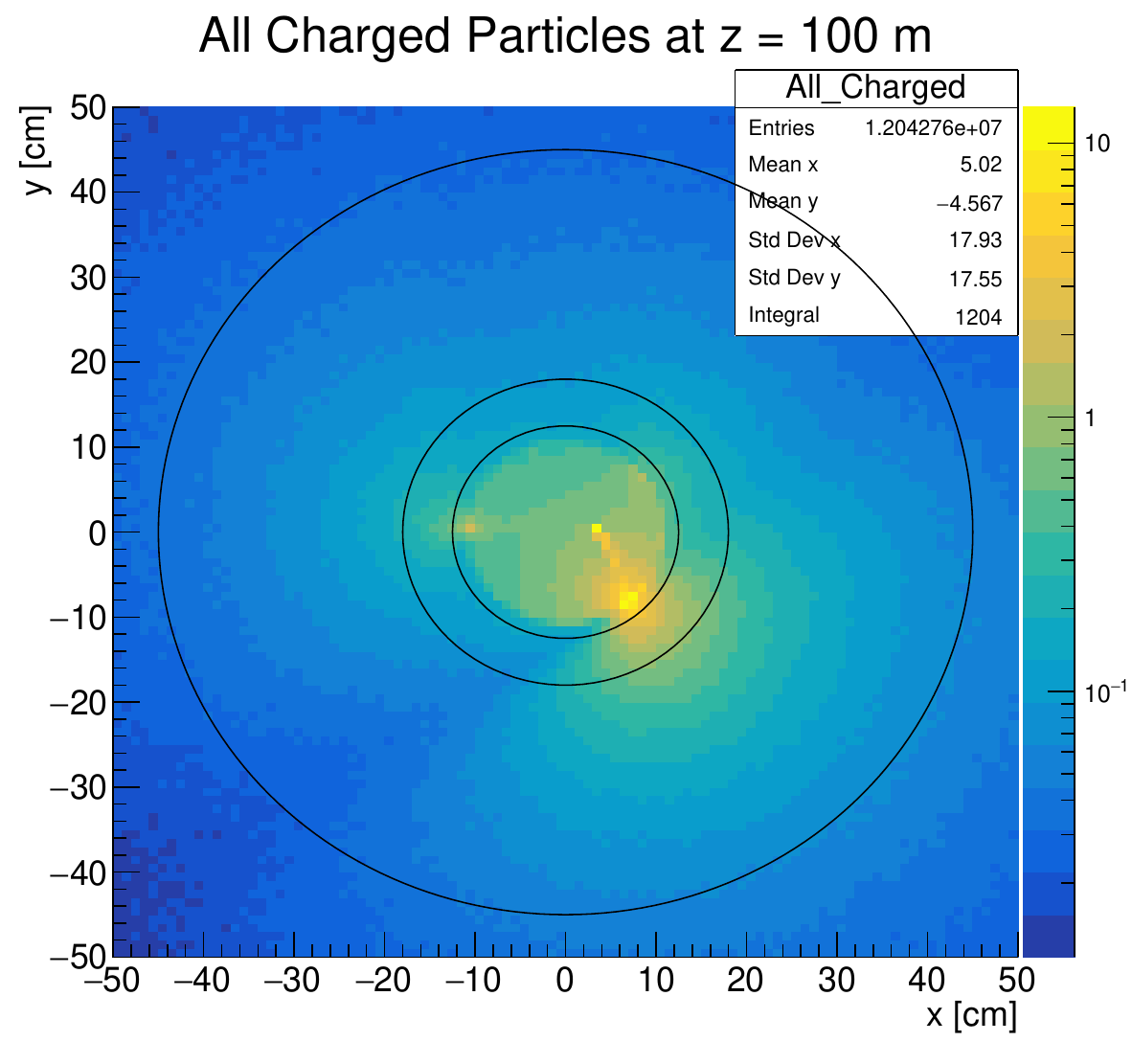}
    \includegraphics[width=0.49\linewidth]{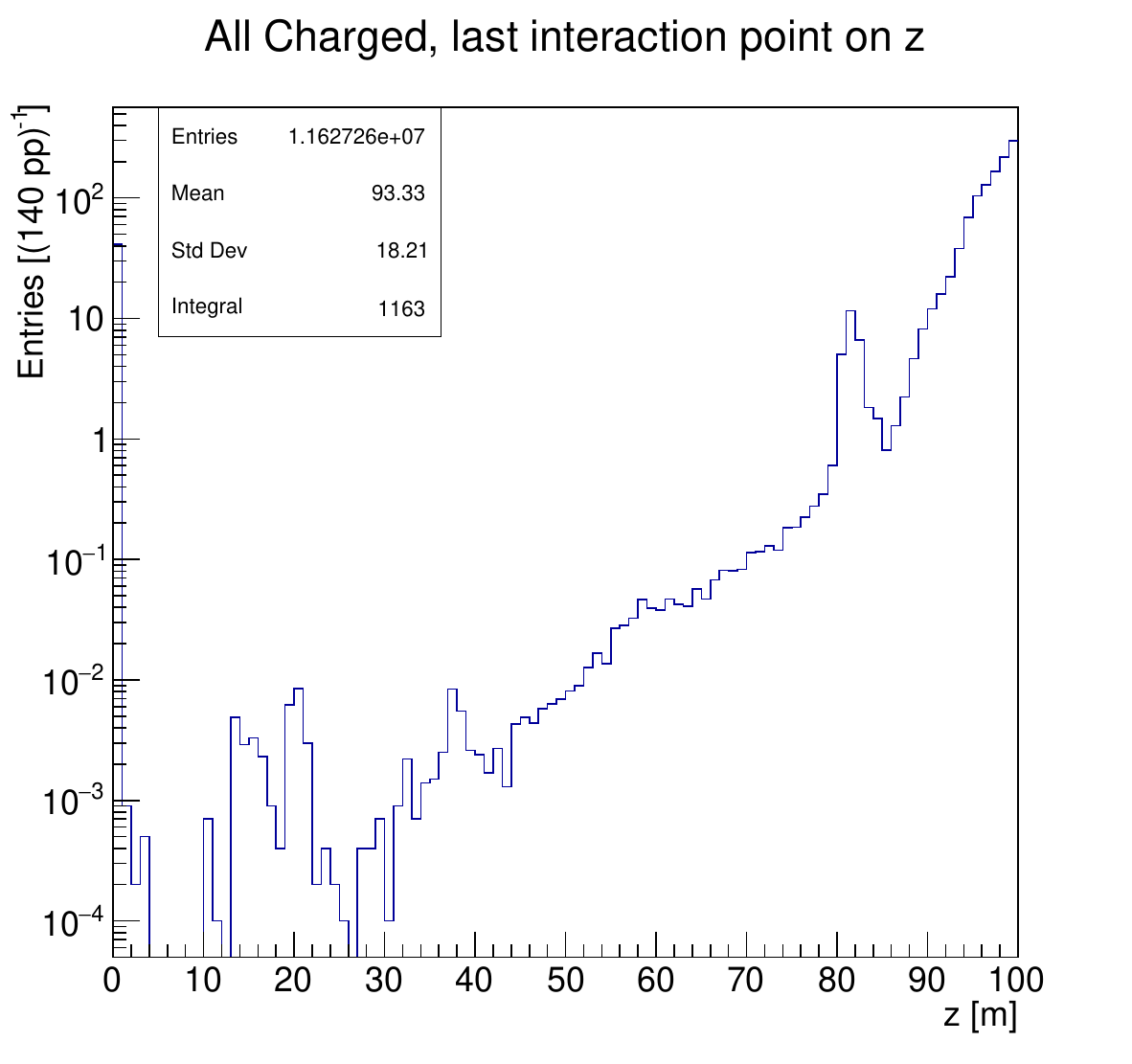}
    \caption{Charged particles produced at IP5 and reaching z = 100 m; xy distribution (left) and z of the last interaction points of those particles (right).  The three circles represent respectively the beam pipe (r = 12.5 cm) after z $\approx$ 108 m, the other two (r = 18 cm, r = 45 cm) are the limits of FACET coverage.}
    \label{fig:All_charged_at_100m}
\end{figure}

In figure~\ref{fig:All_charged_at_100m}, the primary particles from IP5 shown in figure~\ref{fig:ch_neg_and_pos_at_100m} are concentrated inside the pipe (radius 10.6 cm). Secondary particles are mostly produced after 80 m from interactions in the region of D1 and on the downstream pipe. These secondary particles (mostly $\pi^+$, $\pi^-$ and electrons) constitute the bright (yellow) spot near the beam pipe. Detectors near this location will be saturated with a very high particle flux.

\begin{figure}[h]
    \centering
    \includegraphics[width=0.49\linewidth]{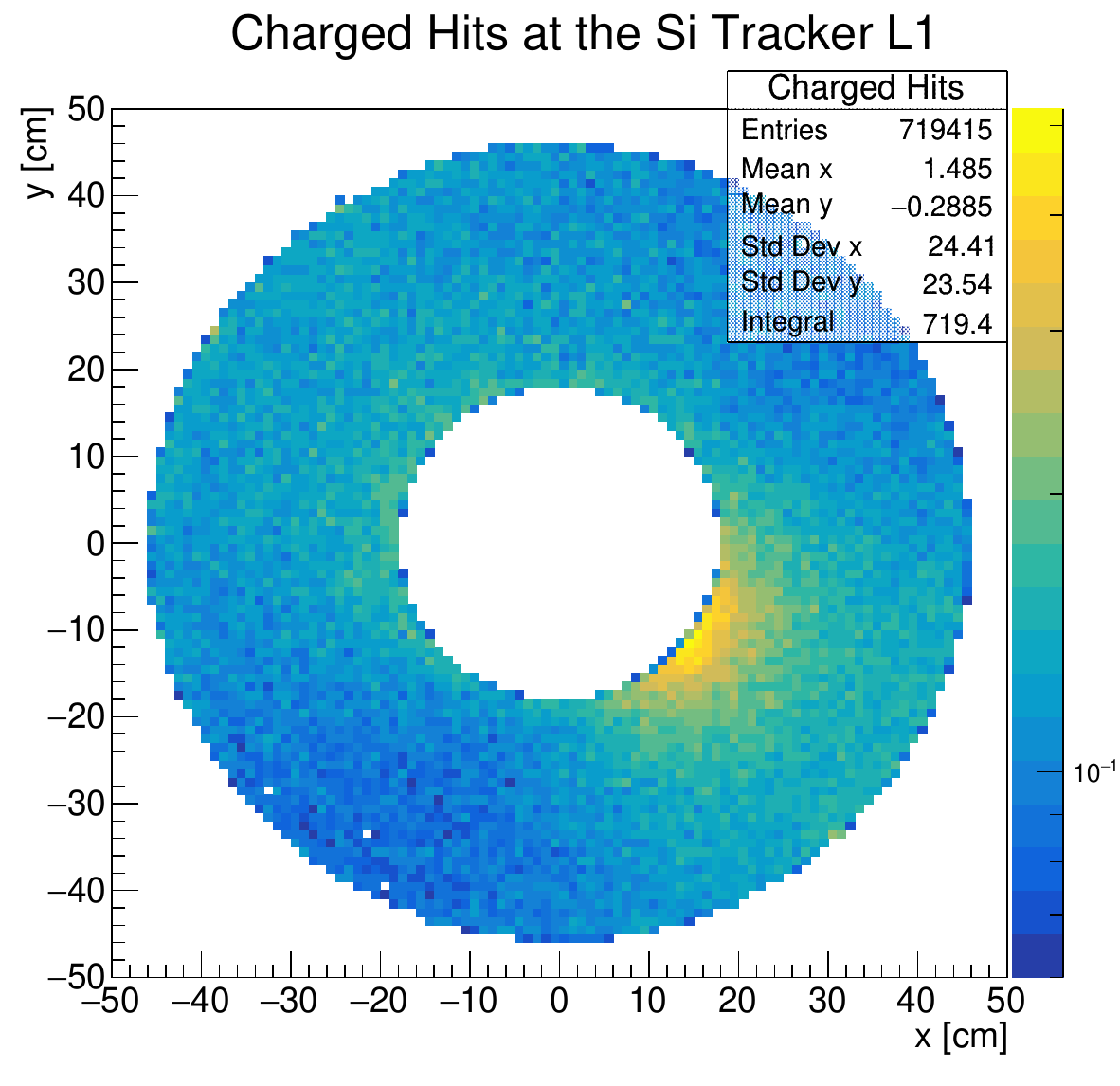}
    \includegraphics[width=0.49\linewidth]{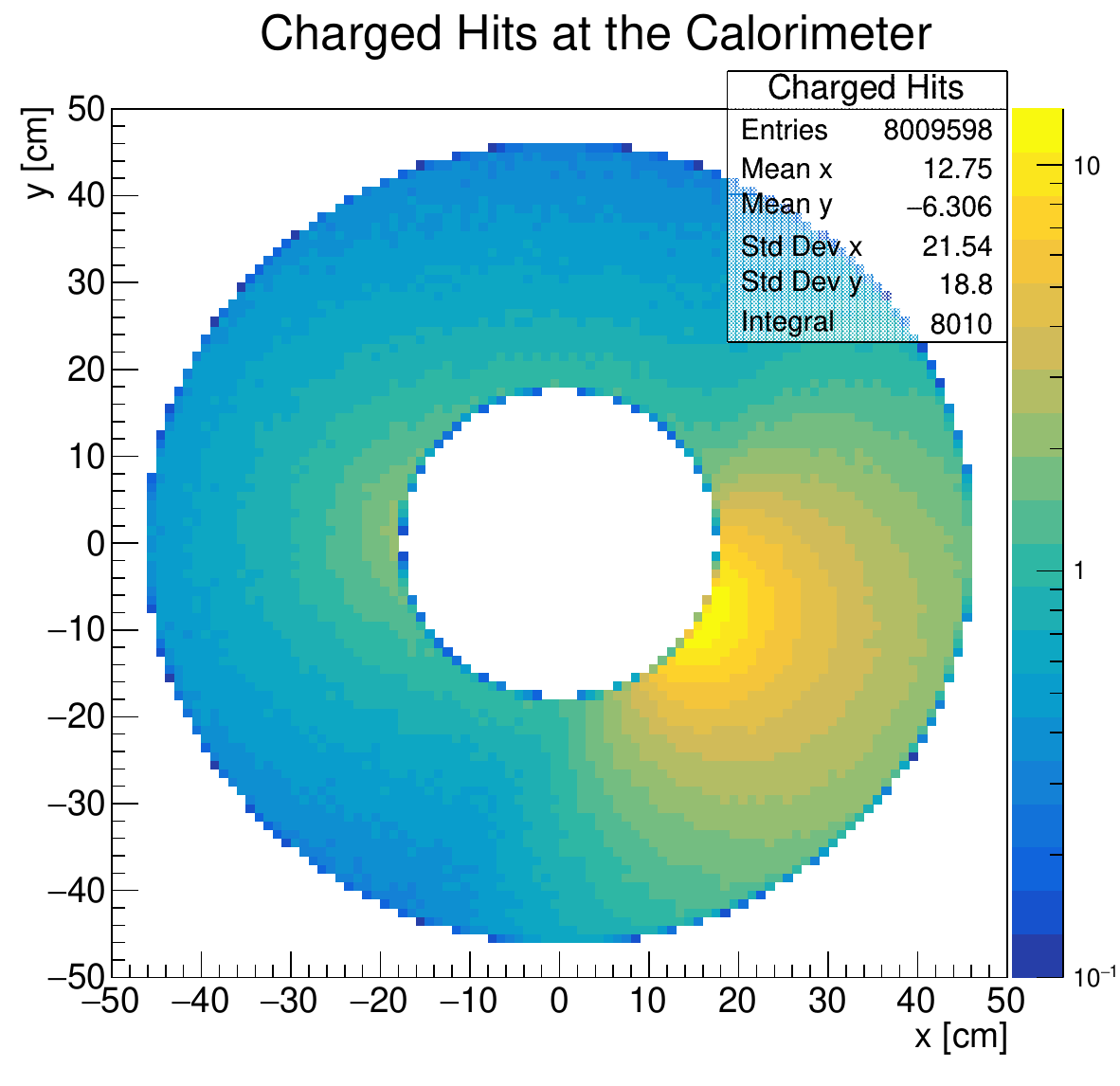}
    \caption{Position distributions (xy) of charged particles (e, $\mu$, pi, p, K) hits at first silicon tracker (left) and the calorimeter (right) are shown for the FACET detector. The beam pipe inside the trackers generates secondary particles. The colored scale shows the flux in (cm)$^{-2}$ (140 pp)$^{-1}$. The maximum flux is lower by an order of magnitude after the calorimeter.}
    \label{fig:FACET_xy_of_charged_at_L1_and_calo}
\end{figure}

Figure~\ref{fig:FACET_xy_of_charged_at_L1_and_calo} presents the xy-spatial distribution of charged particles (e, $\mu$, $\pi$, p, K) at the first silicon tracker (L1) and the calorimeter. Along the trackers, the flux increases from the first tracker to the calorimeter because scatterings between the incoming particles and the beam pipe at the center of the trackers, produce many secondary particles with larger polar angles than the primary ones. Although the calorimeters also have the beam pipe at their center, they exhibit significant absorption, resulting in reduced flux at the sixth tracker (L6), which is located behind the calorimeters.

\begin{figure}[h]
    \centering
    \includegraphics[width=0.48\linewidth]{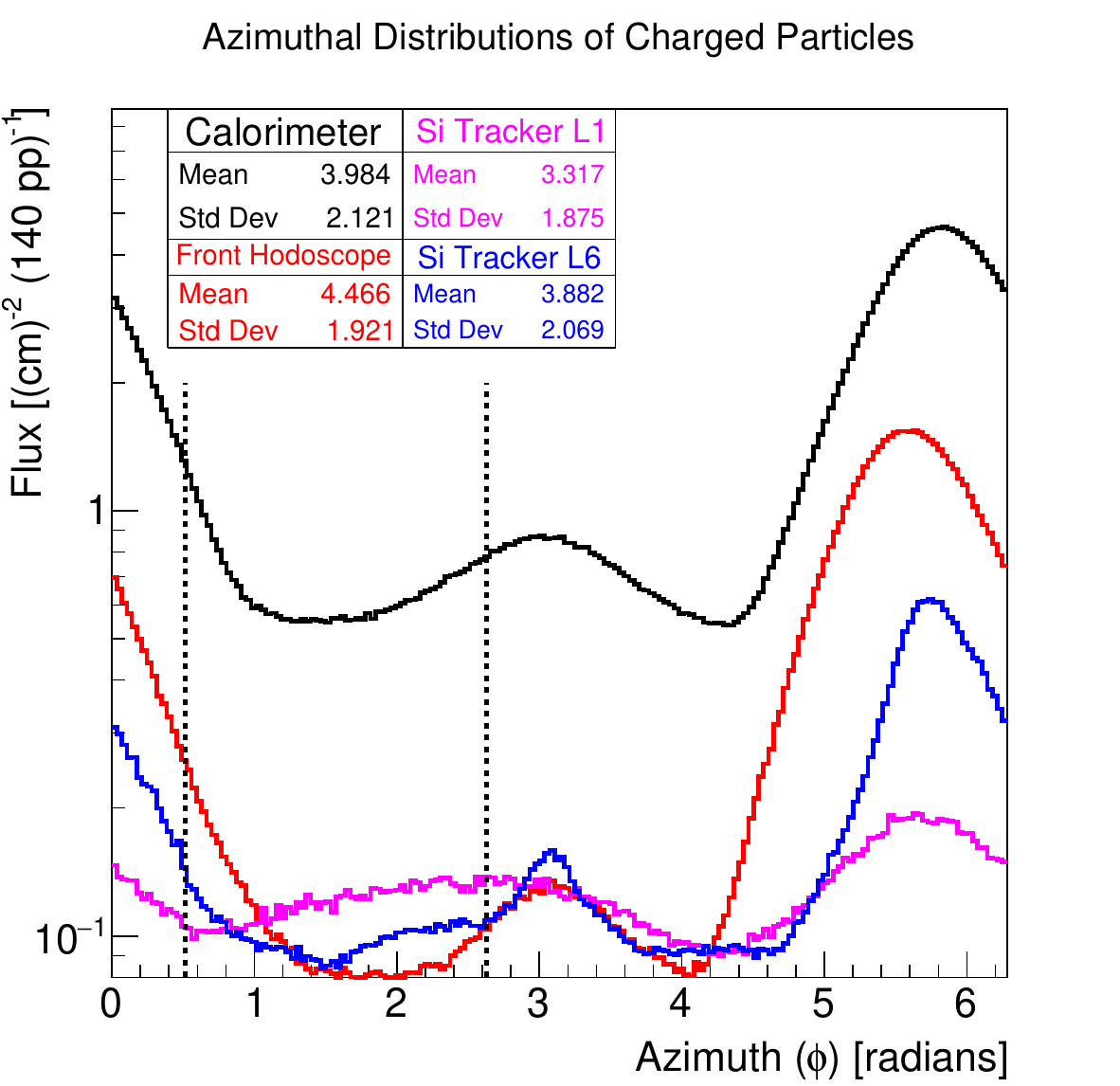}
    \includegraphics[width=0.48\linewidth]{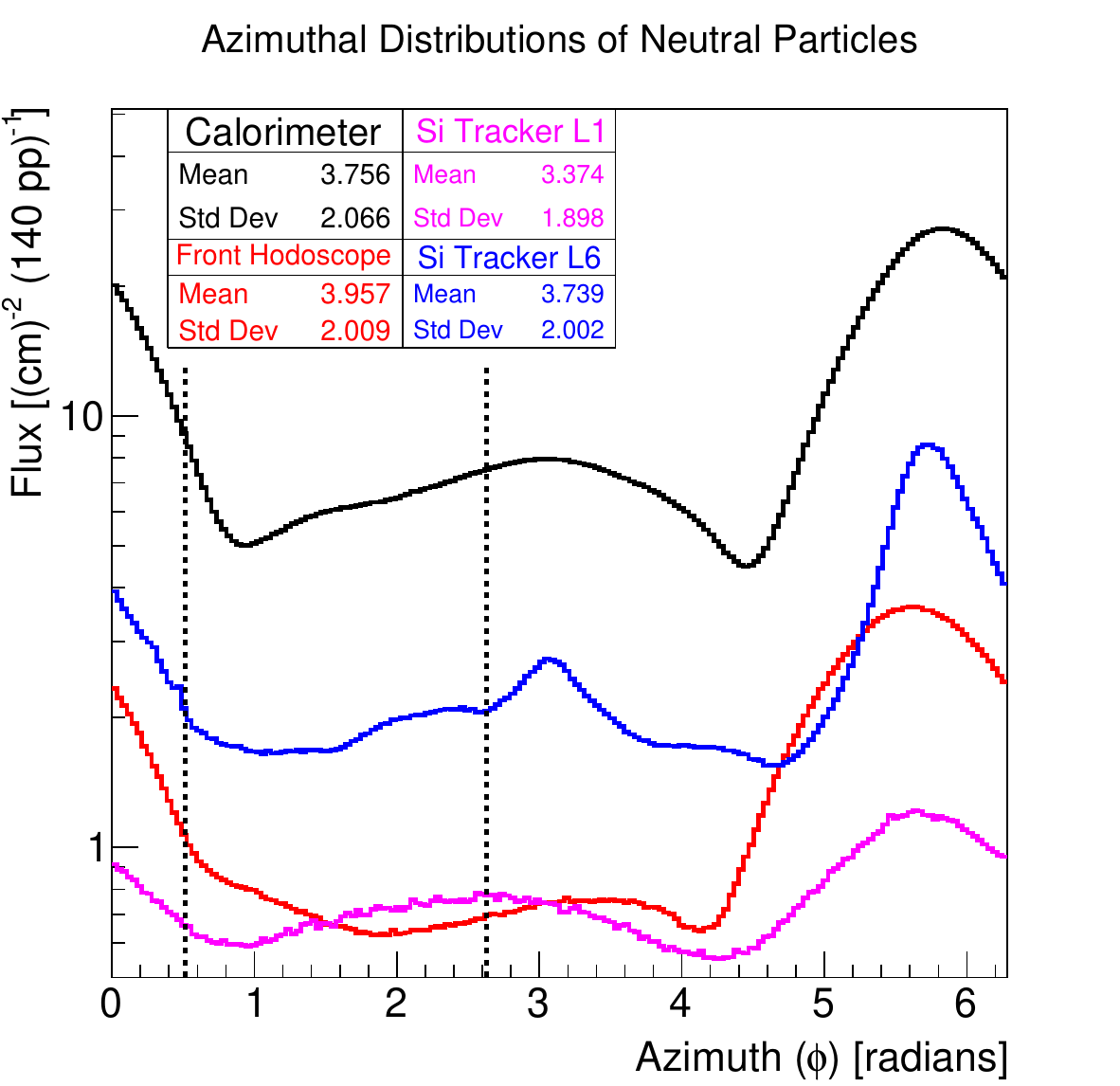}
    \caption{Azimuthal distributions of (left) charged particles (e, $\mu$, pi, p, K) and (right) neutral particles ($\gamma, n, K^{0}_{L}$) hits at different detector layers are shown for FACET detector. The azimuthal region between the dashed lines corresponds to the PREFACE acceptance, excluding the high flux azimuth to the right.}
    \label{fig:FACET_Flux_phi}
\end{figure}

In figure~\ref{fig:FACET_Flux_phi}, the total particle flux on different detectors is distributed over azimuth angle for (left) charged particles and (right) neutral particles ($\gamma$, n, $K_{L}^{0}$). The angle increases in the counterclockwise direction, and shows azimuthal distributions of flux. There are peaks around 6 rads, corresponding to 4 o'clock on the xy-plane (positives) and a second highest flux between 2-4 rads (negatives), corresponding to 9 o'clock on the xy-plane. Dashed lines indicate the azimuthal interval at 12 o'clock, where there is an order of magnitude less flux and where PREFACE would be installed. The full acceptance version of FACET is affected by large particle fluxes, limiting the LLP signal/background; in the restricted azimuthal interval (12 o'clock) the conditions are favourable for PREFACE.

\subsubsection{Limitations imposed for Run 4}

\begin{figure}[h]
    \centering
    \includegraphics[width=1\linewidth]{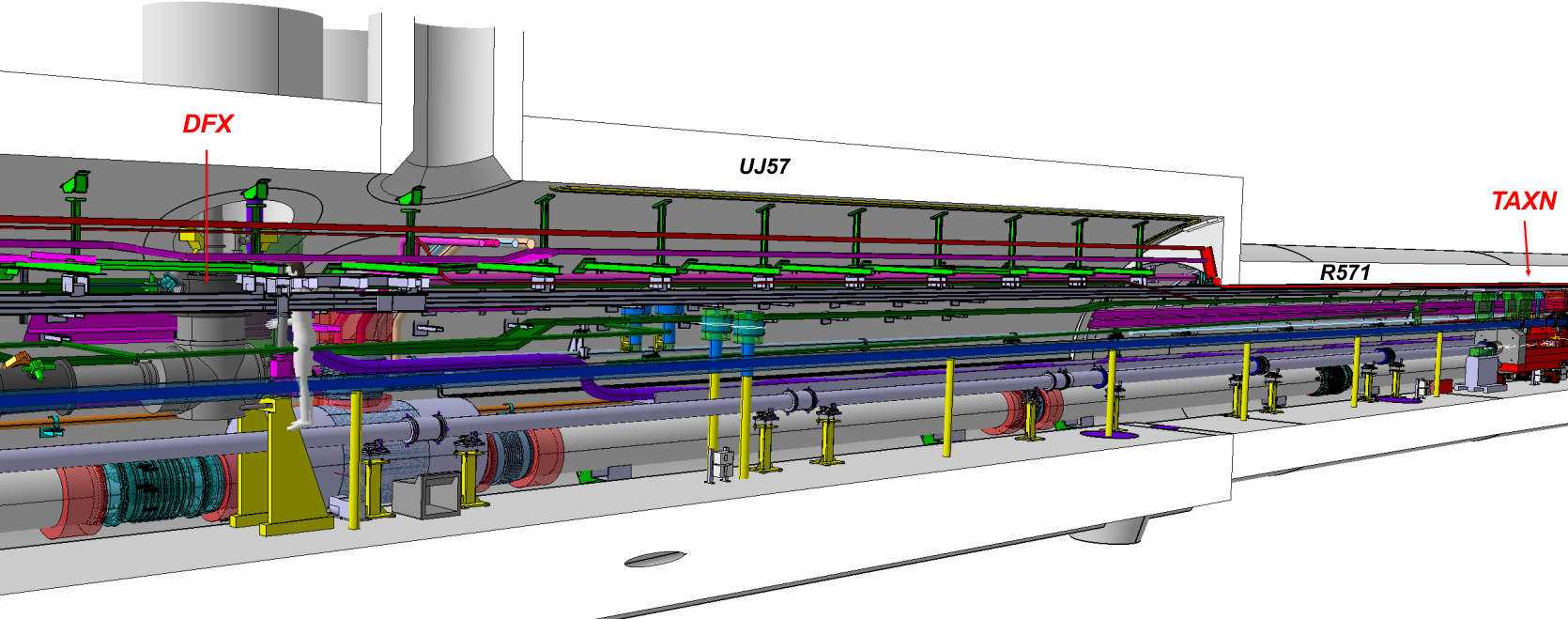}
    \caption{The layout of the LSS5 region between the beam separation dipole D1 and the TAXN absorber~\cite{Fessia2020}.}
    \label{fig:layout_LSS5_D1_to_TAXN}
\end{figure}

\begin{figure}[h]
    \centering
    \includegraphics[width=0.40\linewidth]{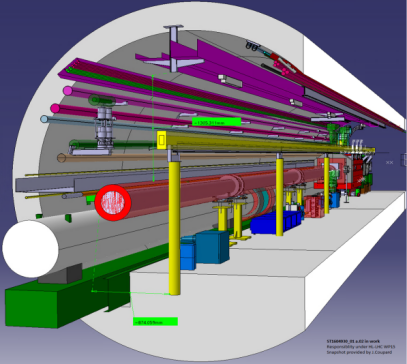}
    \includegraphics[width=0.49\linewidth]{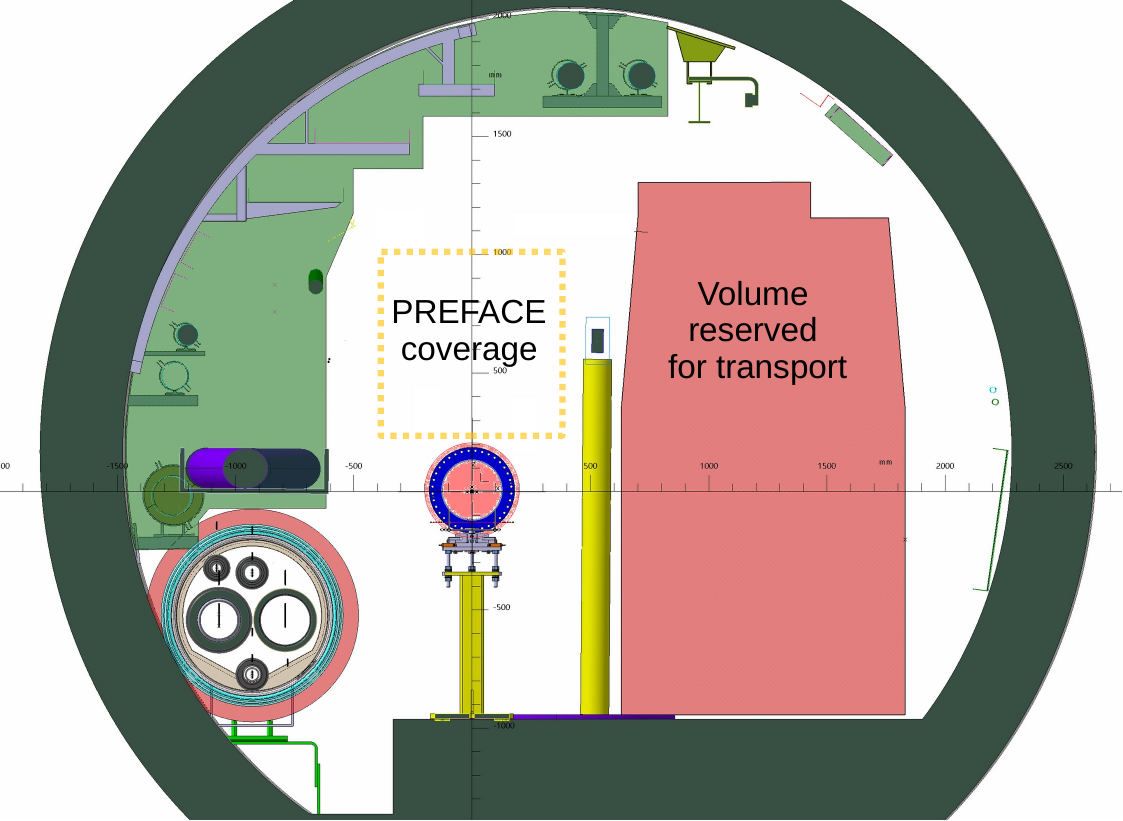}
    \caption{Cross section from 120 m downstream of the IP5, LSS5, LHC tunnel~\cite{Fessia2020}.}
    \label{fig:LHC_xsec_at120m}
\end{figure}

For Run 4 the setup has to be compatible with the presence of the "standard" beam pipe in the D1-TAXN drift region, see figure~\ref{fig:layout_LSS5_D1_to_TAXN} and detectors should cover a region above the beam pipe. In figure~\ref{fig:LHC_xsec_at120m}, a cut through the LHC tunnel at z =120 m, shows the limitations on the sides and below the beam pipe and a possible “free access” region above the pipe; the volume in the region marked in yellow is close to 1 m$^{2}$, and would be equivalent to the cylindrical enlarged beam pipe initially foreseen for FACET. In fact, the region compatible with the Run 4 configuration between D1 and TAXN is almost 1 m vertically, for an area of 0.8 m$^2$; now we use 0.8 x 0.8 m$^2$.

\subsection{Figure of Merit for LLP search apparatus}
\begin{figure}[h]
    \centering
    \includegraphics[width=0.8\linewidth]{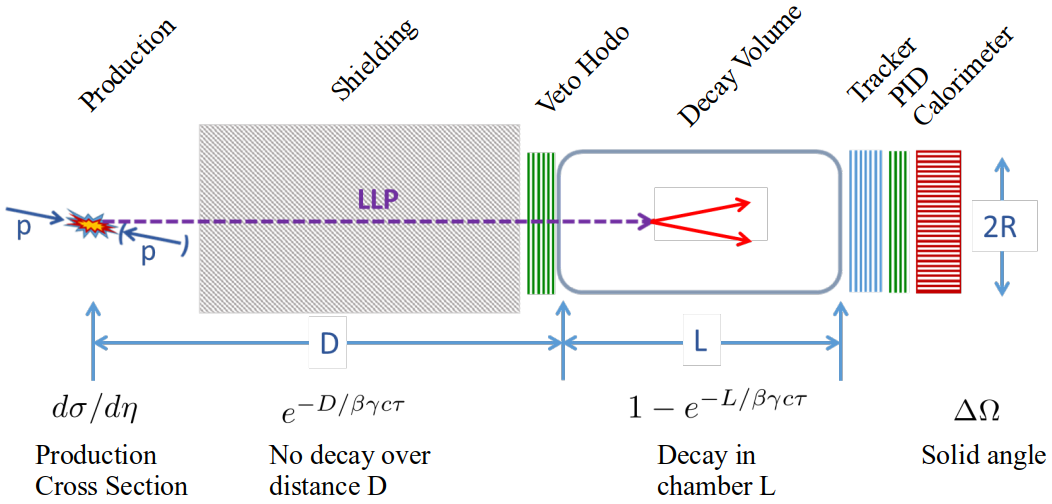}
    \caption{Schematic representation of the parameters defining a Figure of Merit for a forward LLP search experiment.}
    \label{fig:schema_for_FOM}
\end{figure}

In general, an LLP search apparatus can be associated with a Figure-of-Merit (FoM) representing its efficiency over a range of lifetimes (figure~\ref{fig:schema_for_FOM}).

Ignoring the production angular dependence, to be discussed in section~\ref{sec:Physics_case}, the geometric FoM of the apparatus can be described with Equation \ref{eq:FOM}.

\begin{equation}
\label{eq:FOM}
    \text{FoM} = \Delta\Omega\, \exp\left(-\frac{D}{\beta \gamma c \tau}\right) \left(1 - \exp\left(-\frac{L}{\beta \gamma c \tau}\right)\right)
\end{equation}

\begin{figure}[h]
    \centering
    \includegraphics[width=0.65\linewidth]{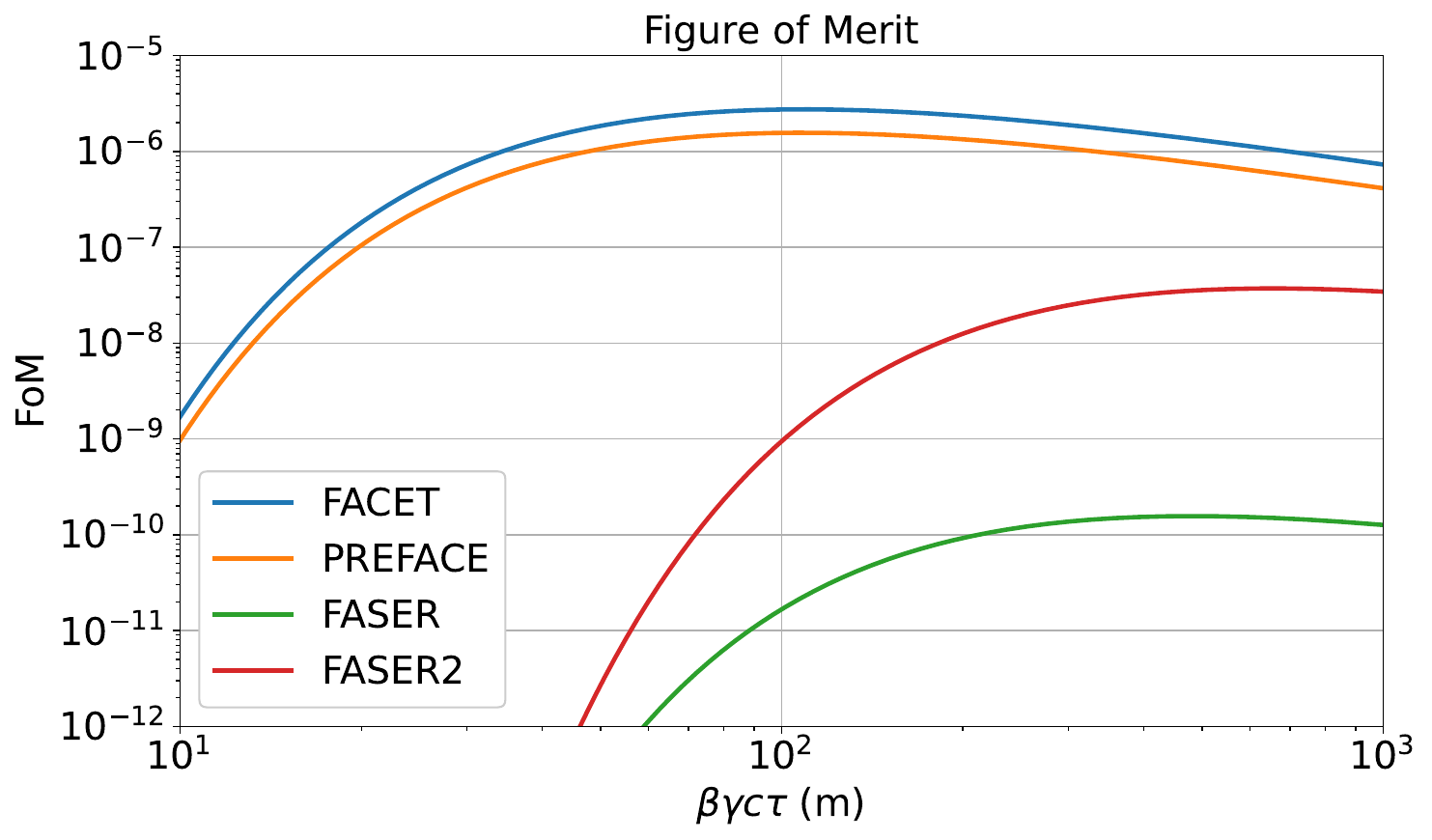}
    \caption{The Figure of Merit for the PREFACE detector and for FACET, FASER and FASER2.}
    \label{fig:FOM at FACET and comparison}
\end{figure}

The geometrical FoM peaks at $\beta \gamma c \tau \approx D$, but extends over a broad interval. In figure~\ref{fig:FOM at FACET and comparison}, the FoM for different LLP projects (table~\ref{tab:experiments}) at the LHC are shown. FACET and PREFACE are similar (having similar solid angles and decay volumes) and are orders of magnitude larger than FASER, which is at a larger distance from the production point, and also compared to FASER2~\cite{osti_1972463}, a project to increase the FASER solid angle. 
For certain channels the angular dependence of the production cross section may favour larger-{$\eta$} setups, as discussed in section~\ref{sec:Physics_case}.

\begin{table}[h]
    \centering
    \begin{tabular}{ccccc}
    \hline
         & Distance & Length of Decay Volume & Geometry & Luminosity\\
        \hline
        FACET & 100 m & 18 m & R = 0.5 m & 3 $\text{ab}^{-1}$ \\
        PREFACE & 100 m  & 10 m & 0.8 m x 0.8 m & 300 $\text{fb}^{-1}$ \\
        FASER & 480 m & 1.5 m & R = 0.1 m & 300 $\text{fb}^{-1}$ \\
        FASER2 & 650 m & 10 m & 3 m x 1 m & 3 $\text{ab}^{-1}$ \\
        \hline
    \end{tabular}
    \caption{Main parameters of the forward LLP experiments: Proposed FACET~\cite{Cerci:2021nlb}, PREFACE, FASER~\cite{Abreu_2024} and FASER2~\cite{osti_1972463}.}
    \label{tab:experiments}
\end{table}

\section{PREFACE setup}

\begin{figure}[h]
    \centering
    \includegraphics[width=1\linewidth]{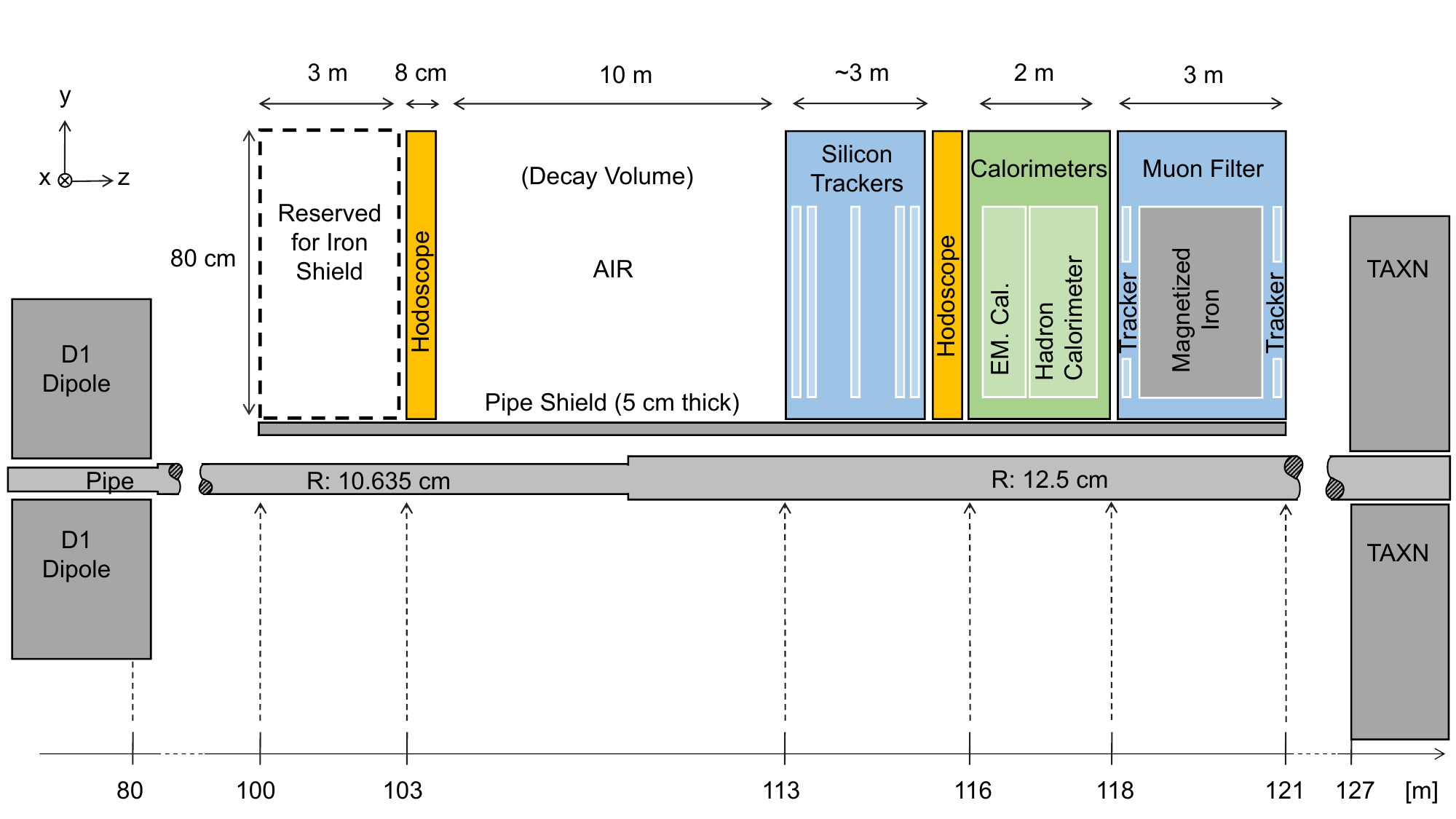}
    \caption{The layout of the \pf setup (not to scale).}
    \label{fig:sketch_preFACET}
\end{figure}

We have investigated a version (figure~\ref{fig:sketch_preFACET}) of the proposed PREFACE detector which is compatible with the standard LHC beam pipe in LSS5, covering only a region above the pipe. PREFACE has the same solid angle as FACET and extends to larger polar angle $\theta$, while avoiding the highest fluxes shown in figure~\ref{fig:ch_neg_and_pos_at_100m}. It is protected from backgrounds from particles interacting in the beam pipe by a shielding plate between the pipe and the detectors. A track-based trigger will select events with two or more tracks on a common vertex in a fiducial decay volume. LLPs produced at the IP5, having small interaction cross sections, will penetrate 35 - 50 m of magnetized iron in the LHC quadrupole and dipole magnets and can decay to SM particles in the PREFACE detector.

\subsection{PREFACE geometry}

\begin{figure}[h]
    \centering
    \includegraphics[width=1\linewidth]{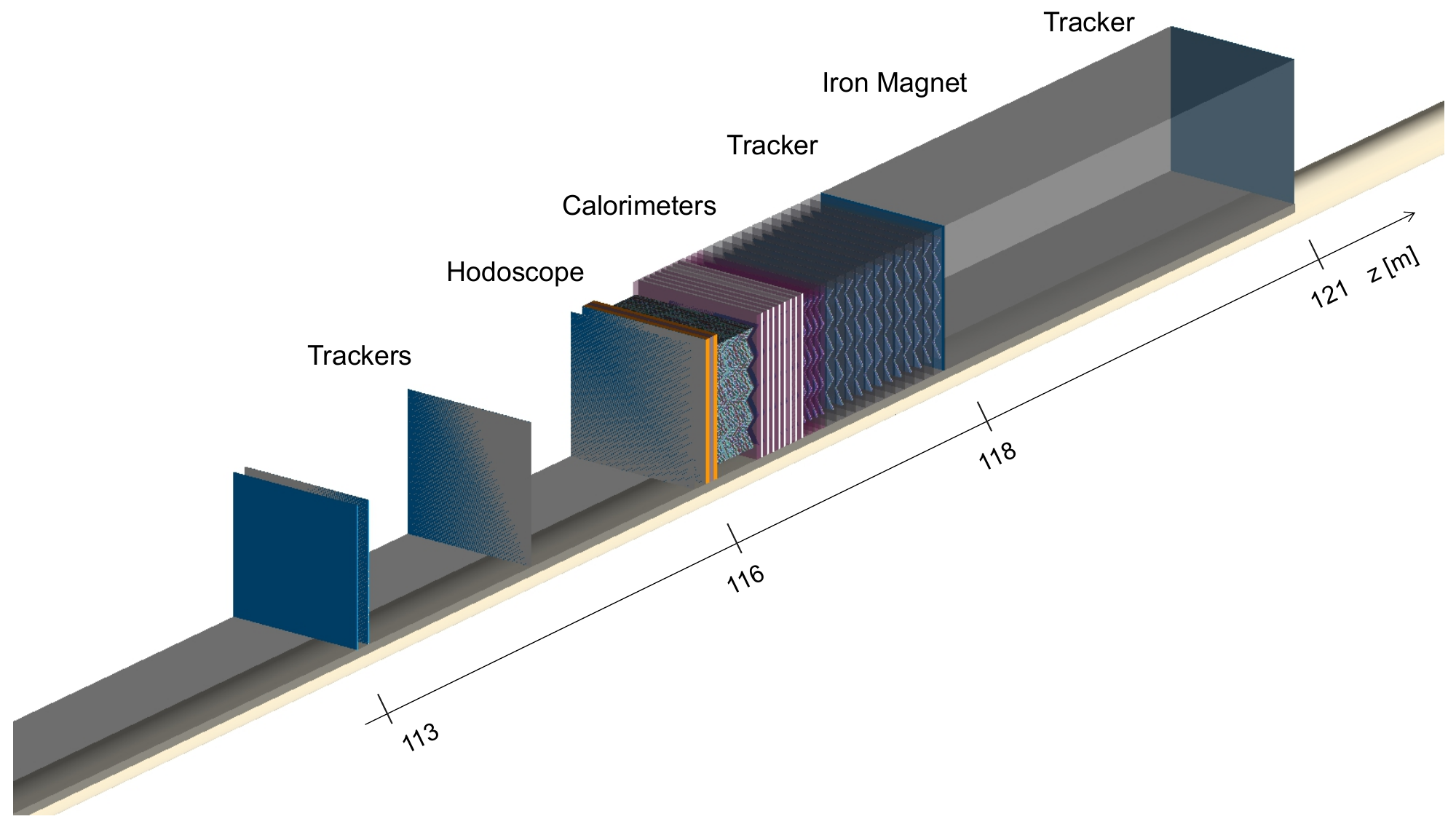}
    \caption{The geometry of detectors downstream of the decay volume for the \textsc{Geant4} PREFACE simulation.}
    \label{fig:down_detectors_from_Geant4_PREFACE}
\end{figure}

\textsc{Geant4} PREFACE simulations have been performed. Input particles reaching a scoring plain at z = 100 m are obtained from a \textsc{Fluka} simulation compatible with the Run 4 LSS5 configuration. \textsc{Geant4} PREFACE simulations start with a 3 m shield block, followed by the front hodoscope/TOF system and tracker, decay space and downstream detector array. Additionally a muon identifier is defined after the calorimeters; it contains a (possibly permanently magnetized) iron shield between trackers. There is also pipe shield beneath the detectors. In figure~\ref{fig:down_detectors_from_Geant4_PREFACE}, \textsc{Geant4} geometry of the PREFACE detector is shown. Detectors are located above the pipe and pipe shield.

\subsection{Background estimates}

\begin{figure}[h]
    \centering
    \includegraphics[width=1\linewidth]{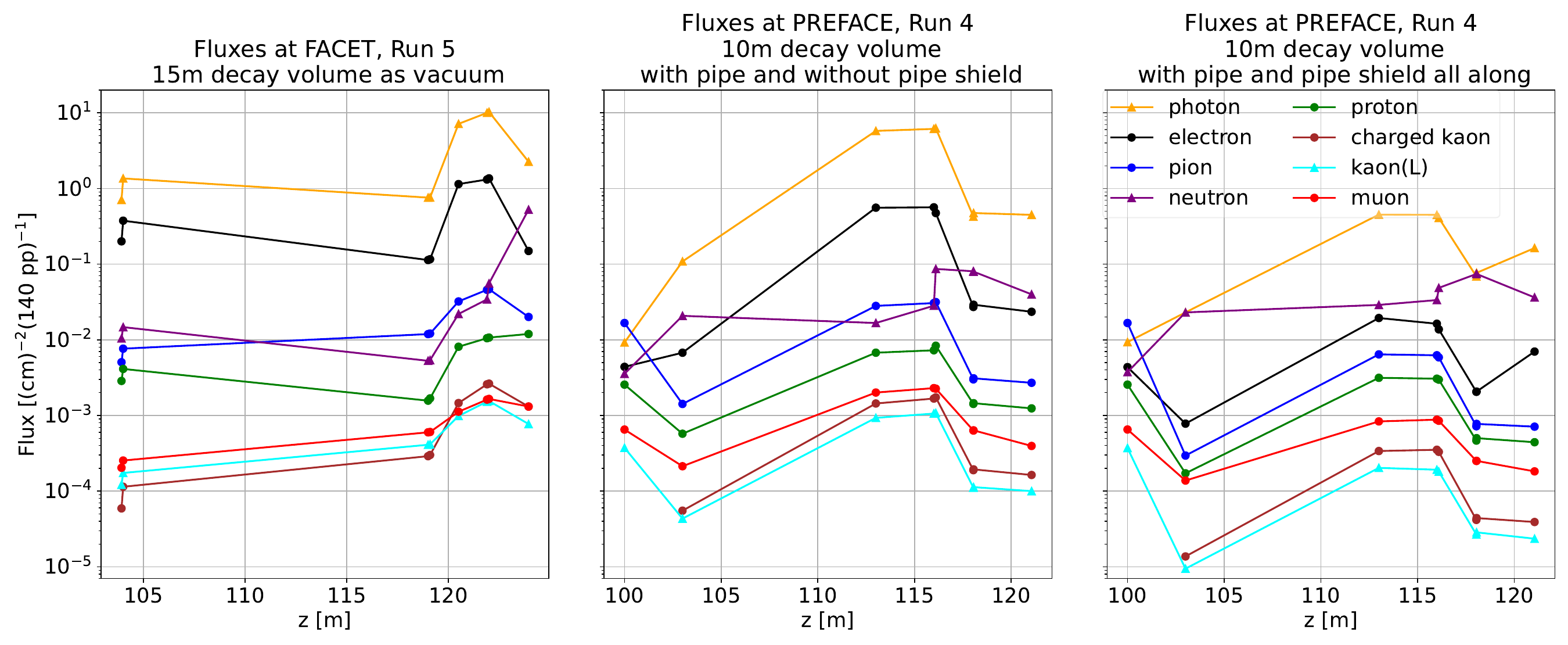}
    \caption{The flux of different particle types comparing three conditions: (left) FACET, (center) \pf without pipe shield and (right) \pf with pipe shield. From center to right plot, the reduction of background is an order of magnitude.}
    \label{fig:bkg_facet_comp}
\end{figure}

\textsc{Geant4} simulations give detailed information on backgrounds. In figure~\ref{fig:bkg_facet_comp}, flux distributions over $z$ positions are shown for 3 different setups. FACET (left) is compared with PREFACE variations: without pipe shield (center) and with pipe shield (right). Hits of charged particles ($e, \mu, \pi, p, K$) and neutral particles ($\gamma, n, K^{0}_{L}$), including anti-particles, are counted on detector components; front hodoscope, trackers, calorimeters and behind. From left to center plot photons, electrons, muons are reduced by 82\%, 83\% and 54\% respectively behind the calorimeters, PREFACE being only above the beam pipe. Adding the pipe shield further reductions are 84\%, 92\%, 60\% respectively. The design of pipe shield, nominally 5 cm of iron, has not been optimised. Globally, comparing the left (FACET) to right plot (PREFACE with pipe shield), reductions are 97\%, 99\%, 82\% for electrons, charged hadrons and muons respectively.

These simulations have shown that background particle fluxes are much reduced from the FACET proposal, thanks to the azimuthal restriction excluding “hot spots” in the horizontal plane (left/right quadrants), the pipe shield and the increased distance from the beam axis. A preliminary study of physics potential with the \pf geometry shows that some of the significant channels explored for the far-forward detectors are accessible to \pf with a sensitivity enhanced by the shorter distance from IP5 and longer fiducial volume with respect to FASER and FASER2, with a FoM more than an order of magnitude larger than FASER2 in the lifetime region of overlap, and greatly extending the lifetime coverage.

\subsection{Radiation levels in the region foreseen for PREFACE}

\begin{figure}[h]
    \centering
    \includegraphics[width=1\linewidth]{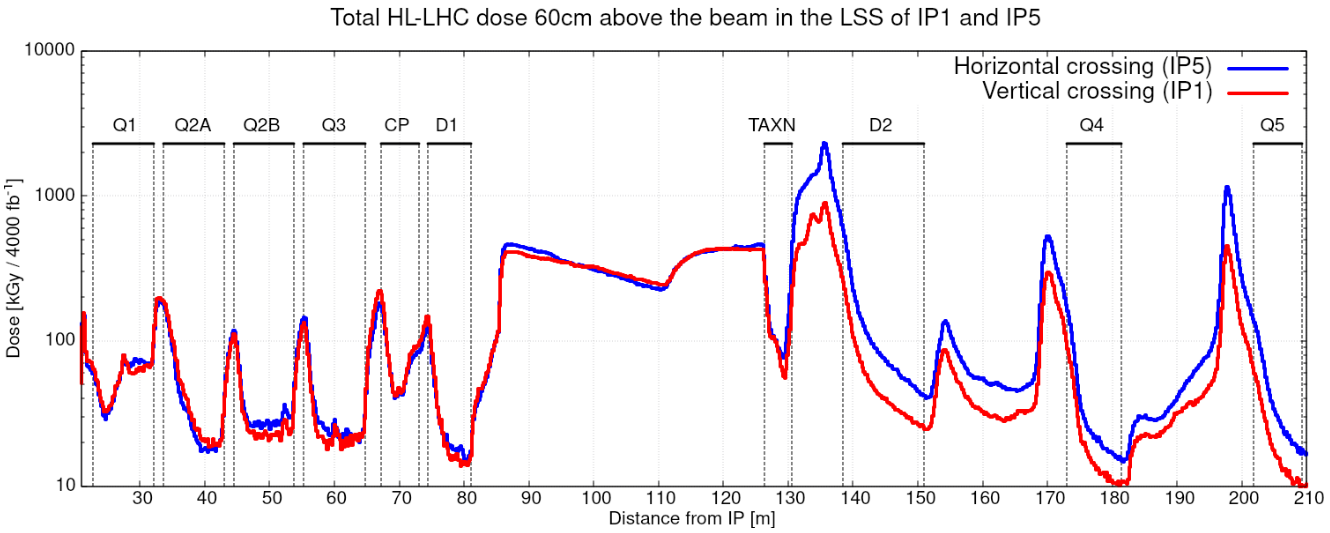}
    \caption{Total Ionizing Dose vs distance from IP5 60 cm above the beamline, estimated with Optics v1.3 of \textsc{Fluka} (A.Tsinganis), for the total HL-LHC run (4000 fb$^{-1}$). The TID for PREFACE ($z$ = 100 m - 120 m) in Run 4 (300 fb$^{-1}$) would be 26 kGy, without the proposed pipe shield.}
    \label{fig:rad_levels}
\end{figure}

\textsc{Fluka} estimates of the radiation levels~\cite{lerner2019hllhc} in the LSS5 region between D1 and TAXN, give Total Ionizing Dose (TID) values at different distances from the IP5, up to 60 cm above the beamline (figure~\ref{fig:rad_levels}), using HL-LHC simulations with optics v1.3 and vertical/horizontal crossing plane in IP1/IP5 respectively.

\section{Physics case for PREFACE}
\label{sec:Physics_case}
The \pf setup has been benchmarked with a few LLP models to assess its physics reach. The models considered are benchmark scenarios recommended by the Physics Beyond Colliders initiative~\cite{Beacham:2019nyx}: Heavy Neutral Leptons (the models BC6-BC8), Higgs-like scalars (the models BC4, BC5), dark photons (BC1), and axion-like particles (BC9-BC11). Their description is given in table~\ref{tab:models}. These models typically have new particles in the GeV range. However, \pf is sensitive to many other models, such as $B-L$ mediators, inelastic dark matter~\cite{Berlin:2018jbm}, and many others, not necessarily limited to the GeV mass range. The sensitivity of PREFACE to these scenarios will be studied later.

\subsection{Comparison to FACET and other LLP projects}
\label{sec:qualitative-analysis}

\begin{table}[h!]
    \centering
    \begin{tabular}{ccc}
    \hline
        Model & (Effective) Lagrangian & Mediator LLP\\
        \hline
        
        HNL $N$ & $\sum_{\alpha}Y_{\alpha}\bar{L}_{\alpha}\tilde{H}N+h.c.$ &
        \begin{tabular}{@{}c@{}} 
        Heavy neutrino $\nu_{\alpha}$ with interaction \\
        suppressed by $U_{\alpha}\sim Y_{\alpha}v_{h}/m_{N}<<1$
        \end{tabular} \\
        
        Higgs-like scalar $S$ & $c_{1}H^{\dagger}HS+c_{2}H^{\dagger}HS^{2}$  &
        \begin{tabular}{@{}c@{}}
        A light Higgs boson with interaction \\
        suppressed by $\sim c_{1}v_{h}/m_{h}$
        \end{tabular} \\
        
        Dark Photon $V$ & $-\frac{\epsilon}{2}F_{\mu\nu}V^{\mu\nu}$ &
        \begin{tabular}{@{}c@{}}
        A massive photon with interaction \\
        suppressed by $\epsilon$
        \end{tabular}\\
        ALP $a$ & $ag_{a}G^{\mu\nu}\tilde{G}_{\mu\nu}+...$ &
        \begin{tabular}{@{}c@{}}
        A $\pi^{0}/\eta/\eta'$-like particle with the interaction \\
        suppressed by $f_{\pi}g_{a}$
        \end{tabular}\\
        \hline
    \end{tabular}
    \caption{Models with long-lived particles considered in this study: Heavy Neutral Leptons $N$, Higgs-like scalars, dark photons, and Axion-Like Particles. Models have variations, and we will study the potential of \f and \pf to particular cases. Namely, HNLs mixing solely with the electron neutrinos (so-called BC6 according to the PBC naming~\cite{Beacham:2019nyx,Antel:2023hkf}), scalars without the coupling $c_{2}$ (BC4) and with $c_{2}$ fixed such that $\text{Br}(h\to SS) = 0.01$ (BC5), dark photons (BC1), and ALPs having non-zero universal coupling to fermions at the scale $\Lambda = 1\text{ TeV}$ (BC10).}
    \label{tab:models}
\end{table}

We now compare the physics reach of \f and \pf 
in terms of the LLP event rates. The main difference is the angular coverage of the detectors. The solid angles are practically the same, $\Omega \approx 4.5 \times 10^{-5}\text{ sr}$. \f covers the full azimuth, $\phi$, but to a more limited polar angle, $\theta$. The impact of this feature can be seen by the solid angle distribution of LLPs, $df/d\Omega \varpropto  df/d\cos(\theta)$. 

Independent of the model, the flux is constant for $\theta < \theta_{\text{peak}}$ and drops at higher angles. The angle $\theta_{\text{peak}}$ depends on the production mechanism. Given the similarity of the total solid angle covered by the detectors, the angular flux of LLPs would be the same at \f and \pf as far as the maximum angle of \pf, $\theta_{\text{max}}$, is less than $\theta_{\text{peak}}$. 

\begin{figure}[H]
    \centering
    \includegraphics[width=0.6\linewidth]{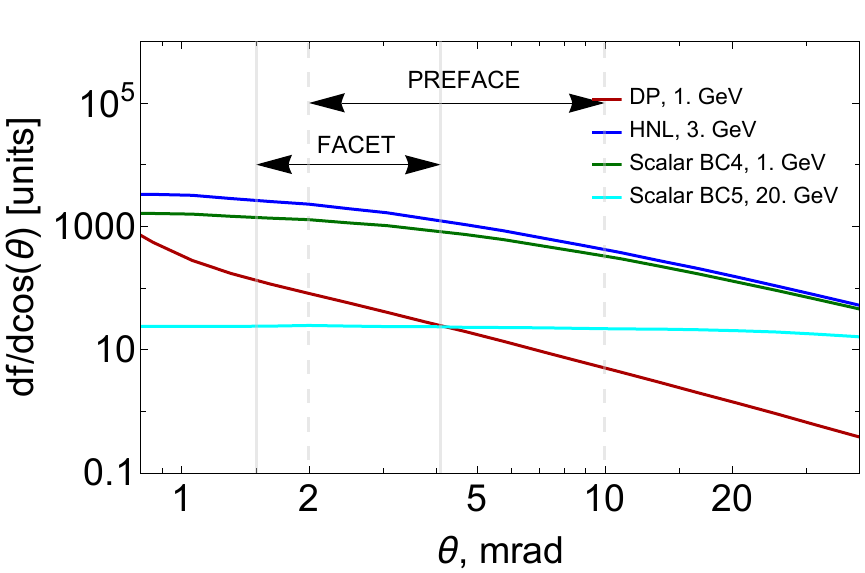}
    \caption{Solid angle distributions $df/d\cos(\theta)$ of various LLPs: dark photons (DP), HNLs, and Higgs-like scalars $S$ with (BC5) and without (BC4) trilinear couplings $hSS$, showing 
 the polar angular coverage of \f and \pf. 
 The plot has been produced using the \textsc{SensCalc} package~\cite{Ovchynnikov:2023cry}.}
    \label{fig:angular_dep_of_proc}
\end{figure}

Figure~\ref{fig:angular_dep_of_proc} shows the angular distribution $df/d\theta$ for a few LLP production processes. For Higgs-like scalars $S$ with an $hSS$ coupling, the distribution is widely isotropic in the angular range of both \f and \pf (explained by the large transverse momentum acquired by scalars in the main production mechanism $h\to SS$). The situation changes for Higgs-like scalars without the $hSS$ coupling, HNLs, and ALPs coupled to fermions. For these, the main production channels are decays of $B$ mesons, and the typical $p_{T}$ values are smaller. The coverage of \f and \pf is slightly above $\theta_{\text{peak}}$. The dark photon case (the production is mainly via proton bremsstrahlung) is characterized by small $\theta_{\text{peak}}$, so the acceptance of \pf is smaller than the acceptance of \f. However, the minimal angle covered by \pf may be smaller than assumed here if optimized shielding allows. 

The large distance of FACET and PREFACE from the interaction point IP5 represents a limitation for LLPs with larger couplings and therefore a smaller lifetimes compared to searches in the main central detectors (the situation in this respect is even worse for FASER and FASER2). These have LLP decay lengths $c\tau\langle \gamma\rangle$  smaller than the distance from the collision point to the decay volume, where $\langle \gamma\rangle$ is the average $\gamma$ factor of the LLP. For the typical energies of LLPs, the decay probability is exponentially suppressed, and only the most energetic LLPs survive to reach the \f and \pf decay volumes. The most energetic LLPs have small $\theta$ values and are more likely to miss the \pf acceptance. This effect is only relevant for the case of dark photons and other vector mediators in the GeV mass range, as we will see in section~\ref{sec:sensitivities}.

\subsection{Sensitivity calculations}
\label{sec:sensitivities}

To calculate the sensitivity of the \pf setup, we use the \textsc{SensCalc} package~\cite{Ovchynnikov:2023cry}. It evaluates the event rate using a semi-analytic approach. It first computes tabulated quantities: (i) the LLP production flux, (ii) the acceptance for it to decay inside the decay volume, and (iii) the acceptance for the decay products. It then calculates the number of events as the integral of the product of these quantities multiplied by the LLP's decay probability. \textsc{SensCalc} has been tested with several event generators -- from light-weight Monte-Carlo simulators to the full experimental framework of SHiP and LHCb. Recently, it has been updated with the \textsc{EventCalc} module, which produces detailed event information using the input from \textsc{SensCalc}, such as tabulated LLP production distributions and decay channels.

\begin{figure}[h!]
    \centering
    \includegraphics[width=0.6\linewidth]{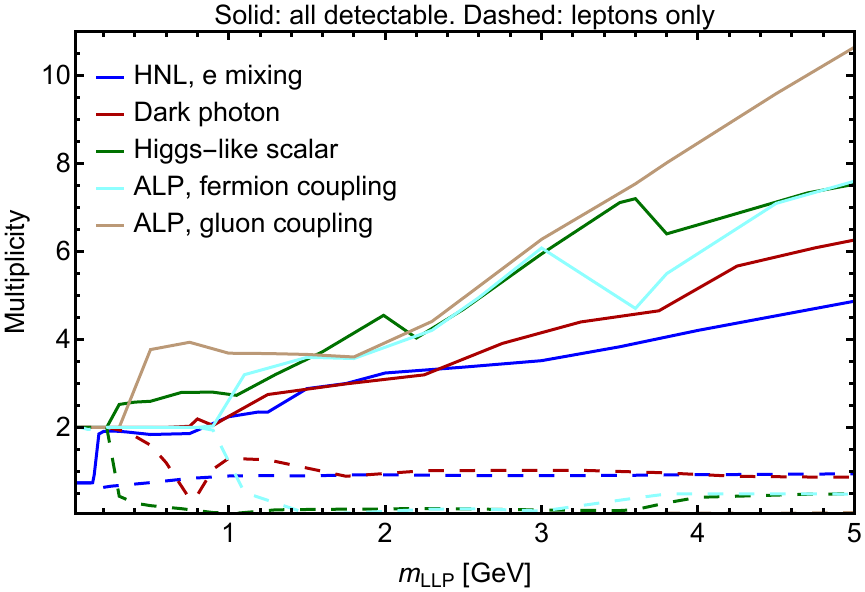}
    \caption{Average particle multiplicity per decay of several LLPs in the GeV mass range, for the models considered in table~\ref{tab:models}. The solid lines show the total particle multiplicity, while the dashed lines show only the charged lepton multiplicity. The figure has been produced using \textsc{SensCalc}~\cite{Ovchynnikov:2023cry}.}
    \label{fig:multiplicity-decays}
\end{figure}

We compute the sensitivity curves of \f and \pf in the plane ``LLP mass $m$'' vs. ``LLP coupling to SM $g$'' requiring $N_{\text{events}}(m,g) \geq 2.3$. That corresponds to the 90\% CL exclusion limit under the assumption of no background. For the detection signature, we consider at least two tracks with a total zero electric charge within the coverage of the last detector plane. Depending on the background situation, other signatures may be used, such as searches for the di-muon vertex or large multiplicity vertices, which is common for LLPs with mass $m\gtrsim 1\text{ GeV}$ when multi-body hadronic decays become possible, see figure~\ref{fig:multiplicity-decays}. We also require the energy of each of the tracks to be $E > 1\text{ GeV}$.
The jumps in the lines are caused either by the behavior of the branching ratios (opening new decay modes, resonant enhancement of some decay probabilities), or by switching from the exclusive description of decays to the perturbative QCD description in the GeV mass range~\cite{Ilten:2018crw,Aloni:2018vki,Bondarenko:2018ptm,Boiarska:2019jym,DallaValleGarcia:2023xhh}. 

When presenting the sensitivity of \pf, in order to compare with other running or future experiments, we consider two plots for each LLP model, see figures~\ref{fig:dp-sens}-\ref{fig:sens-alp}. In the left plots of the figures, we consider near future projects including \pf, FASER, NA62 in the dump mode (called NA62-dump throughout the text), and the downstream algorithm at LHCb~\cite{Gorkavenko:2023nbk}, which we call Downstream@LHCb. We consider the integrated luminosities $\mathcal{L} = 300\text{ fb}^{-1}$ for the first two setups and $\mathcal{L} = 50\text{ fb}^{-1}$ for Downstream@LHCb. For NA62-dump, we assume $10^{18}$ protons-on-target, PoT, corresponding to the luminosity accumulated until Run 3, and use the description of the setup and the selection criteria from the lepton final state analysis from ref.~\cite{NA62:2023nhs}, assuming that the same applies for hadronic states (see the discussion in ref.~\cite{Kyselov:2024dmi}). We take the sensitivity curves of FASER from ref.~\cite{Antel:2023hkf} and for the Downstream@LHCb from ref.~\cite{Gorkavenko:2023nbk}, with the important exception that we use the recent advances in describing the LLP phenomenology instead of the outdated descriptions considered in ref.~\cite{Antel:2023hkf}.

In the right plots of the figures, we show the experiments with the running time extended to the HL-LHC. These are: \f, SHiP, where we assume 15-year running time from 2031 to 2046~\cite{Albanese:2878604}, FASER2 (we take the FPF setup from ref.~\cite{osti_1972463}) and the Downstream@LHCb, where the luminosities correspond to the full high luminosity runs ($\mathcal{L} = 3\text{ ab}^{-1}$ for \f and FASER2, and $300\text{ fb}^{-1}$ for LHCb). We show the NA62 curve assuming $10^{18}\text{ PoT}$.
We use the curves of SHiP from refs.~\cite{Ovchynnikov:2023cry,Albanese:2878604}. We calculate the FASER2 curves using \textsc{SensCalc}, maintaining the same description of the phenomenology used in obtaining the sensitivities.

An important point concerning backgrounds should be taken into consideration. Some of the experiments, like SHiP, have provided detailed background studies showing that they will be background-free, but this is not the case for the other experiments, especially Downstream@LHCb. Therefore, the plots shown below need to be updated once the backgrounds in each experiment are known.

\subsubsection{Heavy Neutral Leptons, HNLs}

Let us start with the case of HNLs. We consider HNLs coupled to electron flavor; the case of other models is similar.

The phenomenology of HNLs is described in ref.~\cite{Bondarenko:2018ptm}. There are three main production mechanisms: decays of $D$ and $B$ mesons, and decays of $W$'s. The hierarchy of these modes is determined by the total amounts of produced mother particles: $N_{D}\gg N_{B}\gg N_{W}$, as well as the kinematic threshold ($D$s may only produce HNLs with mass below $m_{D_{s}}-m_{N}-m_{l}$, where $l$ is the lepton flavor with which the HNLs mix). As a result, decays of $D$ dominate in the mass range $m_{N}\lesssim m_{D}$, $B$ decays in the mass range $m_{D}\lesssim m_{N}\lesssim m_{B_{c}}$, while at larger masses only decays of $W$ contribute. 

\begin{figure}[h!]
    \centering
    \includegraphics[width=0.5\linewidth]{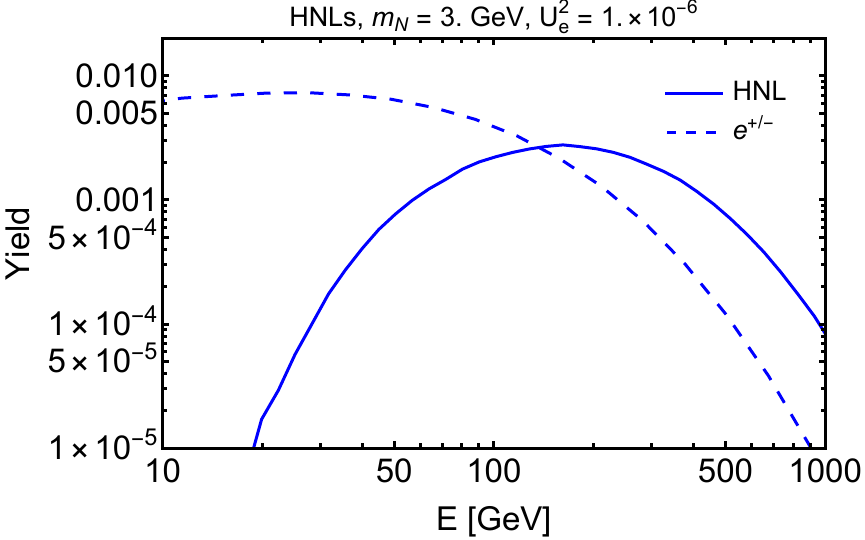}~\includegraphics[width=0.475\linewidth]{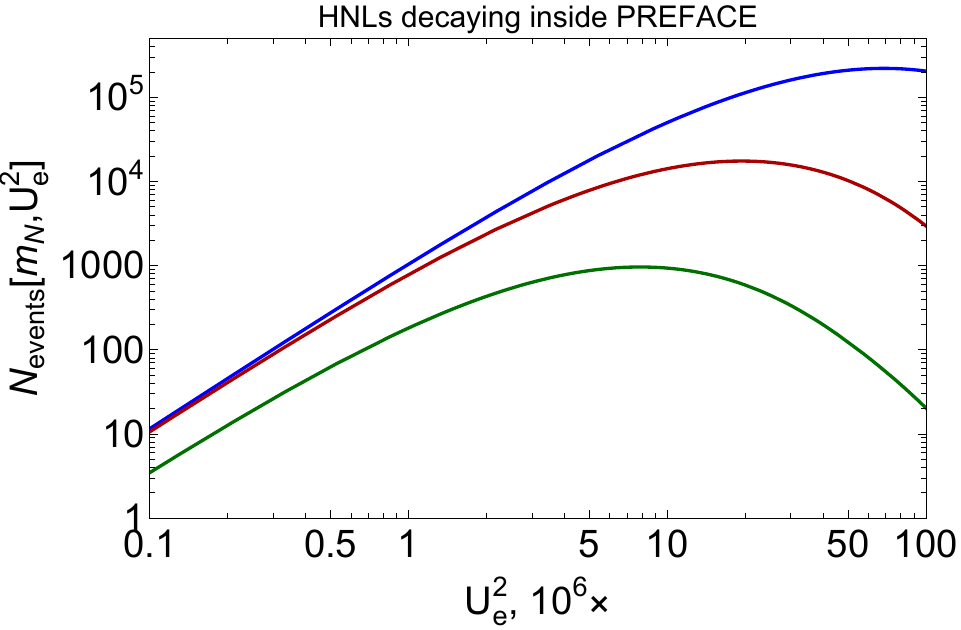}
    \caption{\textit{Left panel:} the energy spectrum of the HNLs decaying inside the decay volume of \pf (the solid blue line) and their decay products $e^{\pm}$ from the decay mode $N\to e^{+}e^{-}\nu$ that point to the end of the decay volume (the dashed blue line). The HNL mass 3 GeV and the squared coupling $U_{e}^{2} = 10^{-6}$ to electron neutrinos are considered as an example. The energy spectrum is normalized to unity. \textit{Right panel:} the number of events in 3 ab$^{-1}$ with decaying HNLs at \pf for the HNL masses $2, 2.5$, and $3$ GeV as a function of $U^{2}$. }
    \label{fig:hnl-spectrum}
\end{figure}

Similar to other LLPs produced in the very forward direction HNLs have mean energies typically a few hundred GeV. The energies of their decay products may exceed a few tens of GeV, see figure~\ref{fig:hnl-spectrum}.

\begin{figure}[h!]
    \centering
    \includegraphics[width=0.5\linewidth]{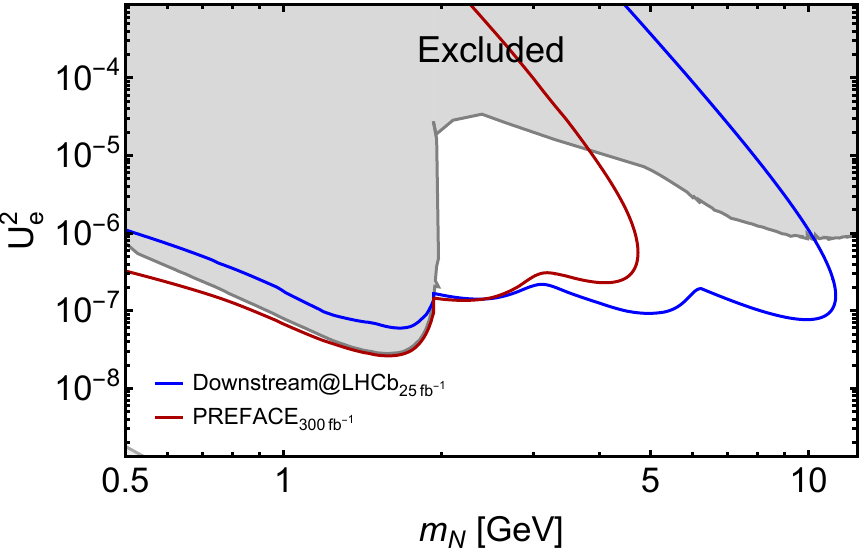}~\includegraphics[width=0.5\linewidth]{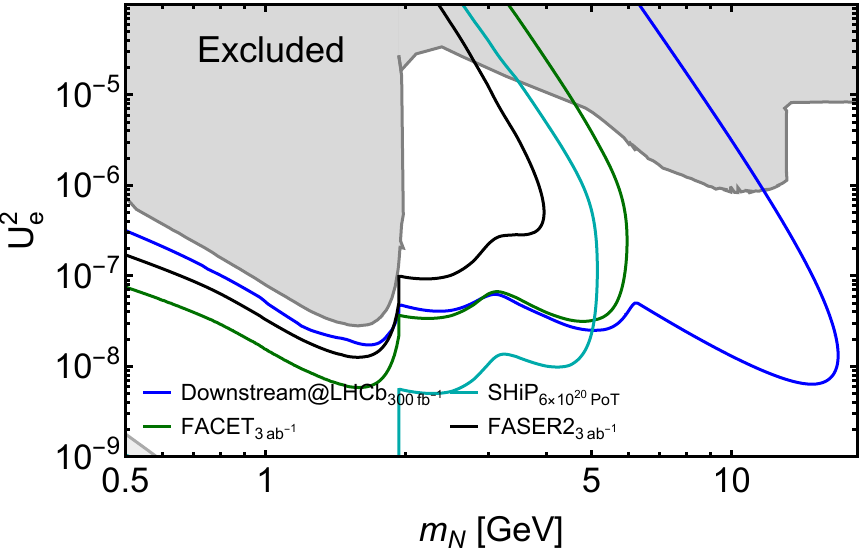}
    \caption{Sensitivities of \f and \pf to HNLs mixing with electron neutrinos (the so-called BC6 model~\cite{Beacham:2019nyx}). In this and other figures below, the left plot covers experiments that could operate in the near future, while the right plot covers the full HL-LHC period.}
    \label{fig:sensitivity-HNL}
\end{figure}

The sensitivity to HNLs is shown in figure~\ref{fig:sensitivity-HNL}. The distribution of HNLs from $D$ is narrower than the distribution of those from $B$, whereas light HNLs from $W$s have the broadest distribution, because of the large $W$ mass. That affects the comparison between the Downstream@LHCb, FASER2, \f and \pf which have the strongest sensitivity among these four experiments for $m_{N}\lesssim m_{B}$ -- they combine the relative far-forward location with a large detector solid angle and a long decay volume. At higher masses, the Downstream@LHCb has a larger solid angle coverage. Together with the much smaller distance to the decay volume with sensitivity to shorter-lived HNLs, it explains why the sensitivity of the Downstream@LHCb algorithm may be extended to the domain above the $B$ meson mass. However, it is important to keep in mind the unknown background status of that search.

\subsubsection{Dark photons}

The phenomenology of dark photons $V$ is described in refs.~\cite{Ilten:2018crw,Foroughi-Abari:2024xlj,Kyselov:2024dmi}. The dominant production modes for $m_{V}\lesssim$ 1 GeV are decays of light pseudoscalar mesons $\pi^{0},\eta,\eta'$, together with primary radiation processes, such as proton bremsstrahlung, final state radiation and production via mixing with vector mesons $\rho^{0},\omega,\phi$, and the Drell-Yan process. 
The decay channels are di-lepton and hadronic decays resembling $\rho^{0}/\omega/\phi$-like modes at masses $m_{V}\lesssim 2\text{ GeV}$, and decays into jets at higher masses.

The proton bremsstrahlung production channel, which is one of the main production mechanisms of dark photons with mass around 1 GeV, suffers from severe theoretical uncertainties~\cite{Foroughi-Abari:2024xlj}. As studied in ref.~\cite{Kyselov:2024dmi}, the sensitivities and constraints on dark photons in the mass range $m_{V}
\gtrsim m_{\pi}$ significantly depend on the bremsstrahlung flux. 
Here, we assume for illustration the ``Baseline'' description from the paper, widely adopted by the community.

\begin{figure}[h!]
    \centering
    \includegraphics[width=0.5\linewidth]{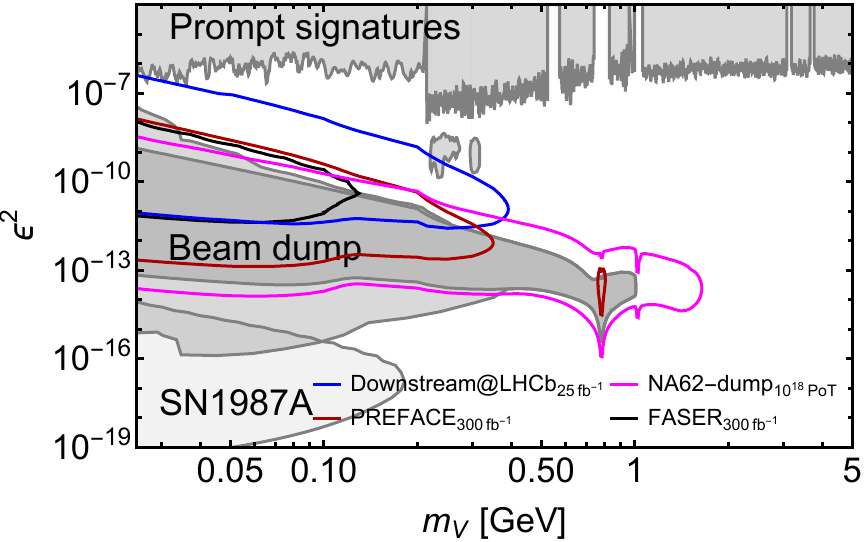}~\includegraphics[width=0.5\linewidth]{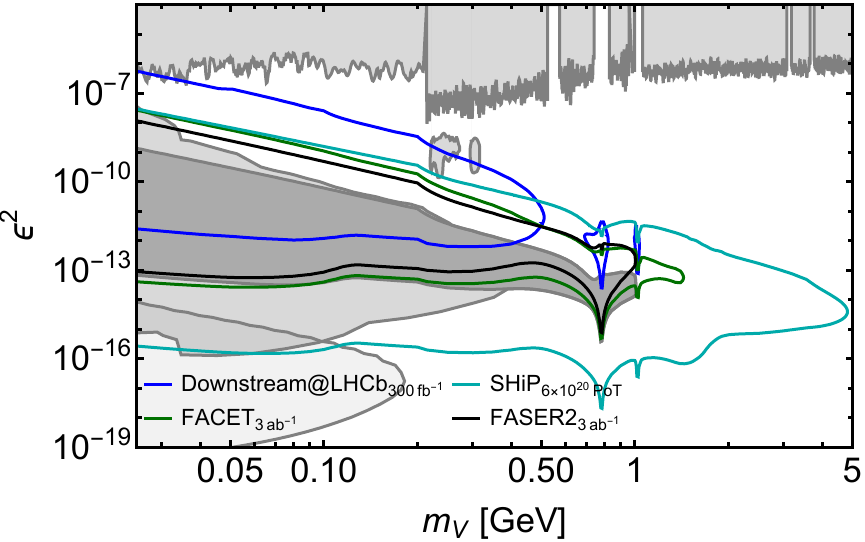}
    \caption{Sensitivities of \pf and \f to dark photons. The constraints on the dark photon parameter space are from refs.~\cite{Kyselov:2024dmi,Antel:2023hkf}. The sensitivities of SHiP, NA62-dump, and Downstream@LHCb have been generated using \textsc{SensCalc}, whereas the sensitivities of FASER and FASER2 have been digitized from ref.~\cite{Antel:2023hkf}. Small islands of sensitivity at $m_{V} \simeq 0.75 \text{ MeV}$ are caused by dark photons mixing with strongly produced $\rho^{0}$ mesons.}
    \label{fig:dp-sens}
\end{figure}

The parameter space of dark photons is shown in figure~\ref{fig:dp-sens}. For the reasons outlined in section~\ref{sec:qualitative-analysis}, both \pf and \f have limited potential to explore the parameter space of dark photons due to the off-axis placement and large distance from the interaction point to the decay volume. Downstream@LHCb has a small distance to the decay volume while FASER is located on-axis, both advantages. Other proposed setups at the LHC, such as SHIFT~\cite{Niedziela:2020kgq}, may have sensitivity to dark photons as well. 

\subsubsection{Higgs-like scalars}

The 125 GeV Higgs $h$ may decay to a pair of light Higgs-like scalars (dark Higgs) $S$ with a small branching fraction. Depending on the presence of the trilinear coupling $hSS$ controlling the decay $h\to SS$, we distinguish between two models: BC4, with $\text{Br}(h\to SS) = 0$, and BC5, where the branching ratio $\text{Br}(h\to SS) = 0.01$ is fixed to be at the boundary of the values accessible by the search for $h\to \text{invisible}$ at the ATLAS and CMS experiments in the high luminosity phase~\cite{Beacham:2019nyx}.

The description of the phenomenology of the scalars may be found, e.g., in~\cite{Boiarska:2019jym}. The main production channel of $S$'s in the BC4 model is the decay of $B$ mesons, originating from the flavor-changing neutral current operator $b\to S+s$. For the BC5 model, the dominant production is the decay of the Higgs bosons, $h\to SS$. The decay modes include di-lepton decays with $\mu^+ \mu^-$ and $\tau^+\tau^-$ favored if kinematically accessible, as well as decays into hadrons. The hadronic decays completely dominate at masses $m_{S}\gtrsim 2m_{\pi}$. In the mass range $m_{S}\lesssim 2\text{ GeV}$, there is significant theoretical uncertainty in the hadronic decays~\cite{Blackstone:2024ouf}, coming from the lack of understanding of the contributions of the intermediate resonances to the di-pion and di-kaon decays. For the given study, we will consider the description from ref.~\cite{Winkler:2018qyg}.

\begin{figure}[h!]
    \centering
    \includegraphics[width=0.5\linewidth]{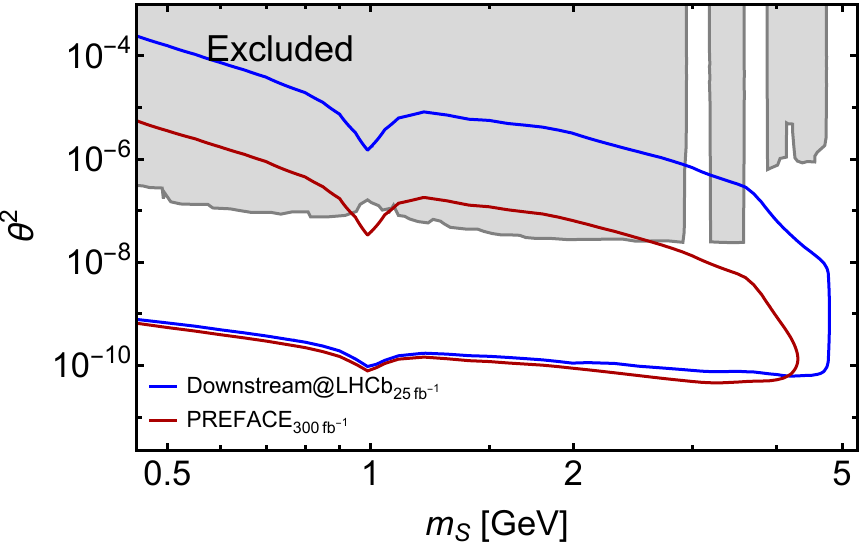}~\includegraphics[width=0.5\linewidth]{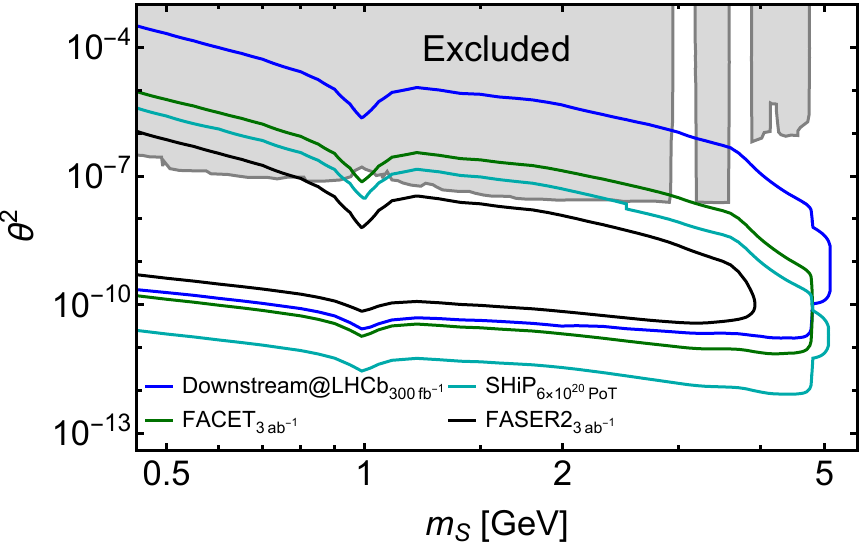}\\ \includegraphics[width=0.5\linewidth]{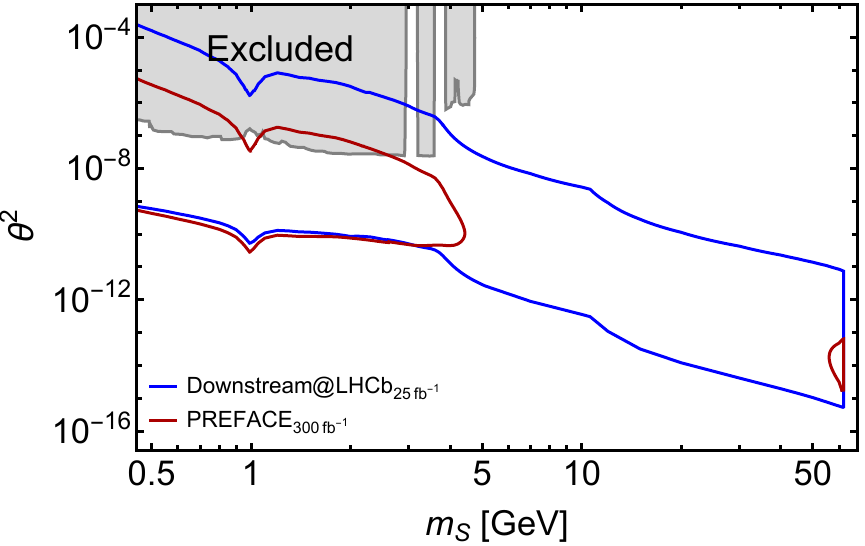}~\includegraphics[width=0.5\linewidth]{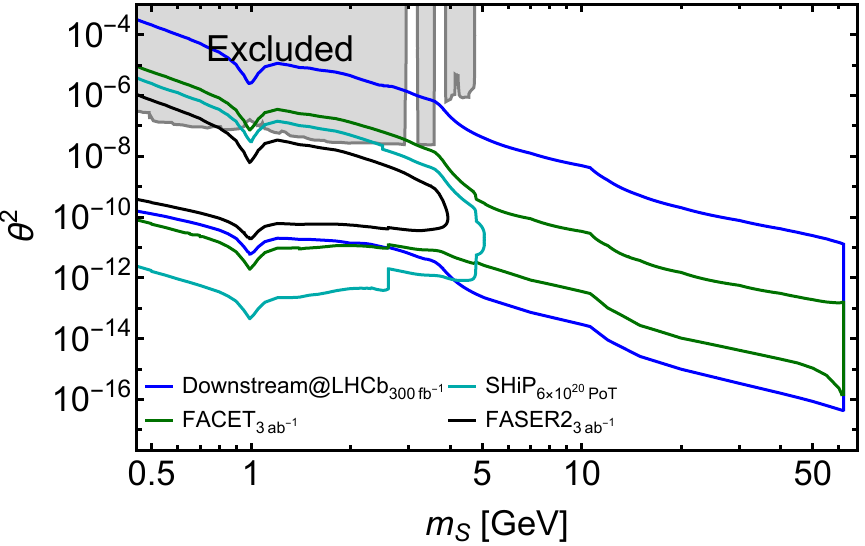}
    \caption{Sensitivities of \f and \pf to Higgs-like scalars: the model without $hSS$ coupling (BC4, top panels), and the model with the $hSS$ coupling fixed such that $\text{Br}(h\to SS) = 0.01$ (the bottom panels). The sensitivity curves have been generated using \textsc{SensCalc} package~\cite{Ovchynnikov:2023cry}. The domain constrained by past experiments has been taken from ref.~\cite{Antel:2023hkf}.}
    \label{fig:sensitivity-scalar}
\end{figure}

The sensitivities are shown in figure~\ref{fig:sensitivity-scalar}. The BC4 case is very similar to that of the ALPs coupled to fermions, as they share a similar production mode -- decays of $B$ mesons -- as well as the Yukawa-like structure of the interactions mediating the decays. We do not show the sensitivity of FASER in the plot with the near-future experiments because it does not extend beyond the region of excluded parameter space~\cite{Antel:2023hkf}. 

For BC5, the production in the decays $h\to SS$ becomes possible. It dominates the yield of produced scalars and makes it possible to probe scalar masses $m_{B}<m_{S}< m_{h}/2$. Scalars from these decays have a broad angular distribution, so the event yield would be larger at the experiments covering large solid angles. FASER2 does not have sensitivity to these masses, whereas Downstream@LHCb provides the best sensitivity if it has a negligible background. 

The sensitivity of \pf with the Run 4 luminosity $\mathcal{L} = 300\text{ fb}^{-1}$ has two disconnected regions -- below $m_{S}\simeq m_{B}$ and at masses $m_{S}\gtrsim 50\text{ GeV}$ because of the mass dependence of the fraction of scalars from the decay $h\to SS$ pointing to the \pf decay volume. If $m_{S}\ll m_{h}/2$, it is very small, of the order of $10^{-4}$. This feature is caused by the large $p_{T}\simeq \sqrt{m_{h}^{2}/4 - m_{S}^{2}}$ gained from the Higgs mass, which makes the angular distribution of scalars more isotropic. Therefore, the main sensitivity comes from the production of $S$ by $B$ mesons. Once $m_{S}$ grows $p_{T}(S)$ decreases and the fraction tends to $10^{-2}$, resulting in more events.

\subsubsection{ALPs coupled to fermions}
One model of axion-like particles $a$ has a universal coupling to SM fermions. The specific choice of the scale $\Lambda$ at which the couplings are defined, $\Lambda = 1\text{ TeV}$, corresponds to the so-called ``BC10'' model under the PBC naming convention~\cite{Beacham:2019nyx}. Its phenomenology has been recently revised in ref.~\cite{DallaValleGarcia:2023xhh}, which considered the missing production modes and hadronic decays of the ALPs.

Similarly to the Higgs-like scalars, the main production channel of such ALPs is the decay of $B$ mesons. The decay modes include di-lepton decays, as well as decays into hadrons, which dominate for ALP masses $m_{a}\gtrsim 1\text{ GeV}$.

\begin{figure}[h!]
    \centering
    \includegraphics[width=0.5\linewidth]{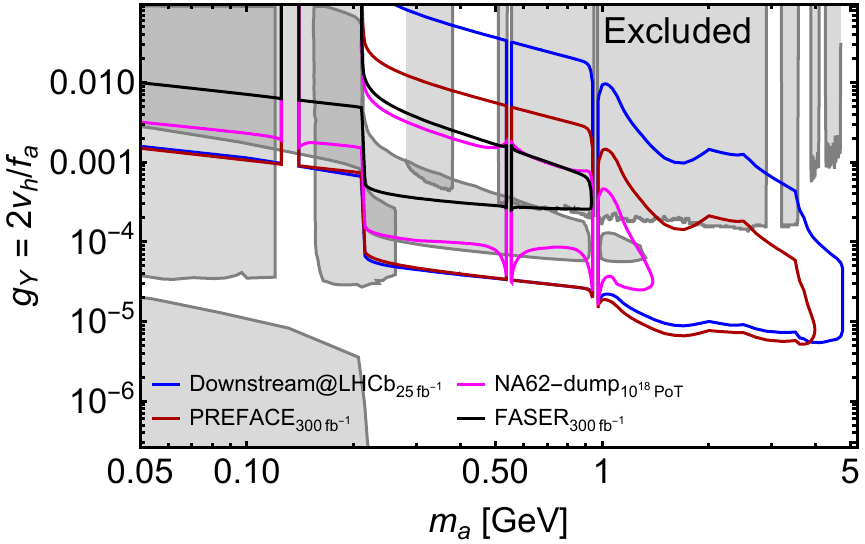}~\includegraphics[width=0.5\linewidth]{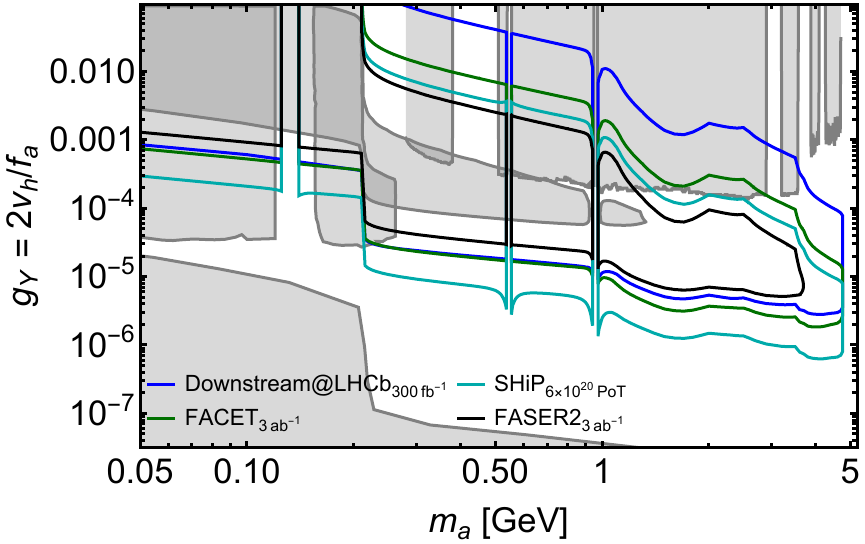}
    \caption{The sensitivity of \f and \pf to the ALPs coupled to fermions in the plane ALP mass $m_{a}$ vs. ALP coupling $g_{Y}$. The description of the ALP phenomenology has been taken from ref.~\cite{DallaValleGarcia:2023xhh}. For consistency between the shown sensitivities, they all have been computed using the \textsc{SensCalc} package~\cite{Ovchynnikov:2023cry}. Note that some of the constrained regions have to be revised to reflect updated phenomenology.}
    \label{fig:sens-alp}
\end{figure}

The parameter space of this model in terms of the ALP mass $m_{a}$ and the ALP coupling $g_{Y}$, together with the sensitivity curves, is shown in figure~\ref{fig:sens-alp}.

The ALPs produced by decays of $B$ mesons have a broad transverse momentum distribution. Therefore, the experiments covering smaller solid angles, such as FASER, NA62-dump, and FASER2 have less acceptance than others covering larger angles such as \pf, and SHiP.

At the lower bound of sensitivity the number of events behaves as $\propto g_{Y}^{4}$, so any factor of 2 difference in the sensitivity curves of various experiments translates to a factor of 16 in the number of events.

At the upper bound of sensitivity, \pf has less sensitivity than the Downstream@LHCb, mainly because of its much farther distance -- the decay volume is located at $z = 100\text{ m}$ compared to $z = 1\text{ m}$ for Downstream@LHCb. At the lower bound, they are similar. Overall, PREFACE provides an excellent exploration potential for ALPs.

In the long-term future, FACET would explore slightly larger parameter space, mainly because of the PREFACE off-axis placement, where the fluxes of many parent particles already start falling (as discussed in section~\ref{sec:qualitative-analysis}). 
PREFACE offers a better sensitivity than SHIP at the upper bound, as the typical momenta of the ALPs are much larger at the LHC. At the lower bound, SHiP would be able to explore much lower couplings, thanks to a much higher luminosity and large angular coverage. These features are largely independent of the model.

\section{An experimental program for PREFACE}

\subsection{Considerations on PREFACE apparatus}
The PREFACE detectors can be optimized considering high resolution tracking (TR) with real time triggering capability, medium-high resolution calorimetry and spectrometry with permanent magnets. The main issues are limited space in the LSS5 insertion region and large backgrounds mainly from secondary interactions in the beam pipe and other materials.

Optimization criteria include a straight path from IP5 to PREFACE between 2 and 10 mrad; the PREFACE apparatus is designed to detect and measure neutral LLP particles produced at IP5, reaching the fiducial decay volume of PREFACE through $\sim 200 \lambda$ of Fe in the quadrupole triplets and dipoles of the LHC lattice. Decays of penetrating particles are measured with the PREFACE detectors behind the decay region.

To remove most background in a prompt trigger, tracks in the first tracker are projected upsteam to the front hodoscope/tracker, and if there is a corresponding hit the track is ignored. If two or more tracks come from a common vertex on the beam pipe walls (pipe interactions) or upstream shielding they are ignored. See details in the trigger criteria section~\ref{subsubsec:Trigger_criteria} below.

\subsubsection{Decay analysis}
For simplicity we consider 2-body decays: $X\rightarrow e^{+}e^{-}, \mu^{+}\mu^{-}, h^{+}h^{-}$. The invariant mass of the 2 SM particles system is

\begin{equation}
    M^{2}_{x} \approx m^{2}_{1} + m^{2}_{2} + 2 E_{1} E_{2} (1-cos(\theta)).
\end{equation}

To accurately measure M$_{x}$, we need good dE/E (the momenta p are only measured for muons) and d$\theta$;
with a (conservative) space resolution for tracker (TR: $\delta$x $\approx$ 200 $\mu$m; l $\approx$ 1 m), $\delta\theta\approx0.2$ mrad; for $X\rightarrow e^{+}e^{-}$, the good angular resolution is complemented by an energy resolution ($\delta E/E\approx0.15/\sqrt{E}\pm 0.01$) for showers in the EM calorimeter; for hadrons ($X\rightarrow h^{+}h^{-}$), the energy resolution would be $\delta$E/E $\approx$ 0.45/$\sqrt{E}\pm$ 0.02 (with e.g. CALICE EM and HAD calorimeter prototypes~\cite{Yoshioka:2013mva}), better if HGCAL modules are available. The HAD resolution would be poor at low energies, but it increases with energy ($\delta$E/E $\approx$ 45\% at 1 GeV/c; at 100 GeV/c, dE/E $\approx$ 4.5\%). 

\subsubsection{Muon spectrometer with permanent magnets}
For the muon channel (($X\rightarrow\mu^{+}\mu^{-}$), a relatively easy and cheap solution would be to use permanently magnetized bricks (Sm-Co, Nd-Fe-B, etc.) with the joint functions of bending the muon trajectories and filtering out hadrons and electrons. For a typical B L $\approx$ 1 T.m and $\delta\theta\approx0.2$ mrad, a momentum resolution dp/p) $\approx$ (5.3 $\times10^{-3}$) could be achieved, but the multiple scattering (MS) in the (Fe, X$_{o}$= 1.76 cm) pile would deteriorate the resolution to about 25\%, limiting the mass resolution of an LLP candidate decaying to $\mu^+ \mu^-$.

The nominal design has 2 m of magnetised iron between tracking stations sections of 0.5 m. The iron can be in (e.g.) four block sections of 0.5 m (weight $\sim$ 2.2 tons/block) with thin tracking layers in between for improved muon track reconstruction. The average rms multiple scattering angle of 50 GeV/c muons is approximately 2.8 mrad, compared to a bending angle of 12 mrad, so the charge is well determined. The momentum resolution is only dp/p $\sim$ 25\%, limited by the multiple scattering, not by the spatial resolution of the tracker. The LHC beam is shielded from the small fringe field  by the pipe shield.

\subsubsection{An alternative air-core permanent magnet}

Maintaining the attractive features of permanent magnets (no power and no - or simple - cooling) and providing unobstructed space for the particles to be momentum-analysed, an air-core design of the permanent magnet can be considered, as in other successful experiments where weight, minimal (or zero-) maintenance, high stability and reliability, are at a premium, like in FASER~\cite{Abreu_2024} and others. A Halbach array type of permanent magnet~\cite{Halbach:1979mv,Blumler2023-hs} offers an interesting alternative at relatively moderate cost.

For PREFACE a possibility is to use an air-core permanent magnet, (typically with Bx $\approx$ 0.5 T (vertical bend); aperture Dx $\approx$ 0.6 m, Dy $\approx$ 0.8 m; length L $\approx$ 2 m). This would have BL $\approx$ 1 T.m (providing a bending angle of 3 mrad at 100 GeV/c), with minimal multiple scattering contributions.

Such a solution allows a rearrangement of the apparatus by
moving the magnet forward between the 2 tracking stations
and moving the calorimeters backward,  with an additional absorber and detection stage behind to tag muons.
In this configuration the setup would be optimized for many more benchmark channels, involving $X\rightarrow e^{+}e^{-}, \mu^{+}\mu^{-}$, and $h^{+}h^{-}$, as well as more complicated (multibody) signatures.
Having the magnet before the calorimeters makes it possible to improve the M(X) measurement and exploit their complementary precisions for low and high energies particles.

\subsection{Remarks on the PREFACE setup}

In this paper we have addressed the possibility of installing and operating the \pf apparatus in the region of the LHC tunnel between the D1 and TAXN in the long straight section downstream of Point 5, taking into account the severe constraints on space, services and radiation in that location. We have profited from the ground-breaking work studying the feasibility of the FACET project, designed specifically in this location in order to allow a strong synergy with the CMS experiment in choice of equipment technologies, operating services and technical support, and event selection and physics measurements.

PREFACE stems from an opportunistic acceptance of the FACET scenario anticipated for Run 5 but for Run 4, without the enlarged beam pipe, and avoiding the highest backgrounds in the horizontal plane. To accommodate the beam pipe configuration for Run 4, it is necessary to restrict the coverage to above the pipe. This region is much less affected by charged particles swept horizontally by D1, and allows additional shielding of interactions in the beam pipe. Unlike the full-azimuth coverage of FACET, the PREFACE detectors can be moved vertically if necessary to minimize backgrounds without reducing the solid angle.

The experimental configuration for PREFACE is not finalized, but the \textsc{\textsc{Geant4}} model that is used in this paper is sufficient to address all principal questions concerning compatibility with the severe backgrounds in the LHC tunnel environment in the HL-LHC conditions.
Essential for PREFACE is the requirement to use high-quality tracking with real time selection of tracks for suppressing residual interactions from the beam pipes, for identifying entering charged particles and for reconstructing topologically interesting decay signatures. Interactions (e.g. of neutrons and $K^0$) on air in the fiducial decay volume have a different topology and may be largely rejected by a higher level trigger, possibly using AI.

\subsection{Trigger criteria}
\label{subsubsec:Trigger_criteria}

Since at high luminosity all bunch crossings will have tracks in PREFACE, track reconstruction and selection (with AI assistance) at the trigger level is essential to reject entering charged particles and tracks from interactions on the beam pipe and other materials. 

Using the results from \textsc{\textsc{Geant4}} simulations, some of these criteria may be tested for each bunch crossing (every 25 ns):

\begin{itemize}
    \item The slope of particle tracks is a powerful diagnostic to reject hits on the beam pipes.
    Tracks with high polar angles, tan($\theta$) $>$ 0.1 can be ignored; they are secondary particles, mostly from scattering in the beam pipes and the pipe shield.
 
    \item Any tracks extrapolated upstream to a hit in the front tracker will be ignored;
      
    \item We require $\geq$ 2 of the remaining charged particles to have a vertex in the decay volume.
    
\end{itemize}

\textsc{Fluka} simulations list identified particles at a score plane (100 m from IP5) with their previous history recorded, whether they come from the primary $p\;p$ collision in IP5, or from subsequent interactions. These different categories may be used to test the trigger criteria concerning not only the pipe interactions, but also pointing back to the primary collision.

At the calorimeter, the simulated (background) particles have maximum energies less than 10, 50, 20 and 20 GeV, for electrons, muons, pions and protons respectively. For neutrals, maximum energies are less than 10 GeV and 20 GeV for photons and neutrons respectively. Moderate cuts on the calorimeter signals will help in effectively cutting backgrounds. As discussed above, while a permanent magnet would be useful to identify muons and measure their charge, the momentum resolution is limited by multiple scattering (MS). An air-core permanent magnet may be preferable to measure the momenta of all charged particles in PREFACE.

In LHC Run 4 with $\gtrsim$ 100 inelastic pp collisions producing many background tracks in PREFACE per bunch crossing, triggering on rare LLP decays will be a challenge, but it is not \emph{a priori} excluded.

In the search for $X \rightarrow c + \bar{c}$ as an example, we will develop a fast trigger based on 

\begin{enumerate}
    \item $\geq$ 1 high momentum muon tracks and/or $\geq$ 1 high energy electron showers with small angles with respect to the beams (``high" means $\gtrsim$ 100 GeV but tuneable), 
    \item At least two tracks in the tracker, that were not detected in the front tracker at the beginning of the decay region, with a distance of closest approach $\lesssim 200 \, \mu$m at a point inside the fiducial decay region, and
    \item a neural net (machine learning/AI) rejecting most interactions on the beam-pipe or shielding while selecting decay candidates.
\end{enumerate}

The Level-1 trigger must give an acceptable rate without dead time, and at a higher level (or off-line) LLP candidates can be selected. Note that there are \emph{no} SM neutral particles with mass above 1.2 GeV/c$^2$ that can decay to charged particles in the fiducial volume, and those with lower mass ($K^0, \Lambda^0$) are produced in shielding, typically with low momentum.

\subsection{Tracking, timing and calorimetry}
From the above discussions and the examples below, it is clear that PREFACE relies essentially on a refined real-time tracking capability, as foreseen for HL-LHC e.g. by CMS. Other fundamental functions such as timing for particle identification and calorimetry, as emphasized earlier, would greatly profit from state-of-the-art technologies, such as LGAD and Si-calorimetry, parts of the CMS upgrades. The number of components for such systems in PREFACE corresponds to a few percent of the CMS upgrade detectors.

\subsection{A search for $X \rightarrow \tau^{+}\tau^{-}, \: c + \bar{c}, \: b + \bar{b}$}
A trigger able to select events with $\geq 2$ tracks including one or more electrons or muons, coming from a common vertex in the fiducial decay volume, enables a search for LLPs in the mass range $4 \lesssim M \lesssim 20$ GeV. These are preferred decay channels for a scalar (dark higgs) in that mass range. The fraction of $\tau$ decays to an electron or muon is 35\%, similar to that of $D^\pm$ mesons. Most pairs of charm and b-mesons have additional charged tracks, so such a trigger can have a high efficiency. (A full simulation with acceptances is under study.) 

Standard-model particle decays from $K^0$ and strange baryons have lower masses and are easily distinguished, and are useful for calibration of mass scales and for testing particle-type identification. Higher flavour particles and $J/\psi, \: \Upsilon$ etc. are all too short-lived to reach the fiducial volume, but could provide a control sample of $e^+ e^-$ and $\mu^+ \mu^-$ pairs if tracks in the upstream tracker are accepted. Their fluxes, as well as those of $K^0$ and $\Lambda^0$ could also provide useful tests of shower simulation programs like \textsc{Fluka}.

Another background is from interactions of high energy neutrons and $K^0$ in air, since unlike FACET, PREFACE does not have a vacuum decay volume. This background can be reduced with a helium bag or a thin-walled tank with air at a reduced pressure. However, such interactions are distinctly different from LLP decays, and most may be rejected even in a trigger. (This is potentially a promising application of AI.)


%

\section{Conclusions}
A search for new long-lived particles produced in the very forward direction, $1 \gtrsim \theta \gtrsim 8$ mr, at the LHC and penetrating $\gtrsim$ 30 m of iron absorber before decaying to SM particles appears to be feasible, provided several conditions can be met. In particular, portals with mass greater than a few GeV decaying to $\tau^+ \tau^-, c \bar{c}$ and  $b \bar{b}$ and inaccessible at fixed target experiments  may be produced at the LHC, waiting to be discovered. The FACET project requires an expanded beam pipe, a vacuum decay region, which is not feasible before Run~5 (not before 2034). We describe PREFACE which does not require any change to the LHC beam pipe and could in principle be operational in Run 4. The solid angle is the same as that of FACET ($\sim 0.64$ m$^2$) but the detectors are above the beam pipe avoiding the highest background regions to the left and right of the beams. A shielding plate between the beam pipe and the detectors gives a further reduction of background from beam halo particles showering in the pipe.  The minimal set of detectors is a precision tracking system in the magnetic field-free region
able to select and trigger on vertices in the decay region above the pipe. This is followed by a permanent magnet (with no power, cables or cooling needed) to measure and trigger on muons. Adding an electromagnetic calorimeter for electrons and photons is important (typical energies for electrons from LLP decays are much higher than from background showers), and a hadron calorimeter would be very desirable. 

At the front of the fiducial decay region a tracker will tag incoming background tracks, with optionally a fast-timing hodoscope, which can incidentally provide valuable information e.g. for testing \textsc{Fluka} simulations of particle fluxes. 

Some key issues for further study are the radiation levels and background minimization, discrimination at trigger level (AI is potentially a key component) between decays of candidate LLPs, SM decays (e.g. $K^0, \Lambda^0$) and interactions.  PREFACE will be a first feasibility demonstration of using this so far unexploited very forward zone with 80 m $\lesssim z \lesssim$ 120 m at the LHC, while initiating a generic search for portals in a new region of parameter space.

\appendix
\section{Future studies in progress}

    \begin{itemize}
    \item Optimize shielding design to minimize radiation doses and track multiplicity. Investigate available radiation-hard calorimeters and trackers with adequate performance. Design magnet optimizing length to minimize hadron track leakage. Design tracking in front of the fiducial decay region with (optional) timing for particle ID of backgrounds (and testing \textsc{Fluka} simulations).
    
\item Calculate acceptance for $X \rightarrow e^\pm, \mu^\pm$ (+ anything) decays as a function of $M(X), p_T, p_z $.  Calculate muon charge sign assignment and momentum resolution dp/p vs p, and use of calibration channels, e.g.  $ J/\psi \rightarrow e^+e^-, \mu^+\mu^-$ (from shielding).
 
\item Design track-based trigger on vertices. Investigate AI trigger classifying events as $X$ candidates, SM decays, or interactions.  
\item Design supports and infrastructure, including signals and DAQ, cables, etc. People, timeline, costs, etc.
\end{itemize}

\acknowledgments 

PREFACE is based on the FACET project, being a study of the possibility of a version of FACET compatible with the LHC beam pipe and infrastructure already in Run~4, and we acknowledge all members of the FACET group for their contributions. We acknowledge valuable contributions from LHC staff, especially F. Cerutti, J.P  Corso, J. Coupard, P. Fessia, and M. Sabate Gilarte.
This work was supported in part by the US Department of Energy under grant DE-SC0010113. Fermilab is operated by Fermi Forward Discovery Group, LLC under Contract No. 89243024CSC000002 with the U.S. Department of Energy, Office of Science, Office of High Energy Physics. 
This study was partly funded by the Scientific Research Projects Coordination Unit of Istanbul University Project number FUA-2022-39051.




\bibliographystyle{JHEP} %
\bibliography{biblio}

\end{document}